\documentclass{ejm}

\usepackage{hyperref}
\usepackage{amsmath}
\usepackage{amssymb}
\usepackage{tabulary}
\usepackage[pdftex]{graphicx}
\graphicspath{{../figs/}}
\usepackage[round]{natbib} 
\usepackage{subfigure}

\newsavebox{\astrutbox}
\sbox{\astrutbox}{\rule[-5pt]{0pt}{20pt}}

\newtheorem{theorem}{Theorem}[section]
\newdefinition{definition}[theorem]{Definition}
\newtheorem{lemma}{Lemma}

\usepackage{xcolor}

\newcommand\Pain{Painlev\'e }
\newcommand\GB{G\'enot-Brogliato }

\title{Transitions and singularities during slip motion of rigid bodies}

\author[P. L. V\'arkonyi]{%
  P.\ns L.\ns V\ls \'A\ls R\ls K\ls O\ls N\ls Y\ls I$\,^1$,
}

\affiliation{%
  $^1\,$Dept. of Mechanics, Materials and Structures, Budapest University of Technology and Economics, Budapest, Hungary}

\date{\today}
\pubyear{2018}
\volume{000}
\pagerange{\pageref{firstpage}--\pageref{lastpage}}

\begin{document}

\label{firstpage}
\maketitle

\begin{abstract}
The dynamics of moving solids with unilateral contacts are often modeled by assuming rigidity, point contacts, and Coulomb friction. 
The canonical example  of a rigid rod with one endpoint
 slipping in two dimensions along a fixed surface (sometimes referred to as Painlev\'e rod) has been investigated thoroughly by many authors. The generic transitions of that system include three classical transitions (slip-stick, slip reversal, lift-off) as well as a singularity called dynamic jamming, i.e. convergence to a codimension-2 manifold in state space, where rigid body theory breaks down. The goal of this paper is to identify similar singularities arising in systems with multiple point contacts, and in a broader setting to make initial steps towards a comprehensive list of generic transitions from slip motion to other types of dynamics. We show that - in addition to the classical transitions - dynamic jamming remains a generic phenomenon. We also find new forms of singularity and solution indeterminacy, as well as generic routes from sliding to self-excited microscopic or macroscopic oscillations.
\end{abstract}

\begin{keywords}
contact dynamics; friction; rigid body dynamics; singularity; piecewise smooth systems.
AMS subject classification: 37J55,
70F40,
70E18,
70K70,
70F35
\end{keywords}

\section{Introduction}
 Slipping contact between solids is responsible for various important dynamic phenomena, such as dynamic jamming, friction-induced vibrations, brake squeal, stick-slip and sprag-slip oscillations. For many engineering applications, it is crucial to understand and predict the onset of these phenomena, as they are associated with impulsive contact forces, mechanical damage, unwanted noise and intensive wear of the interacting surfaces \citep{ibrahim1994friction,sinou2007mode,butlin2013friction,mo2013effect,le2013friction,kruse2015influence}. 
 
Moving solids with contact interactions are often modelled as rigid multi-body systems with unilateral point contacts subject to Coulomb friction. Even within this simple framework, it is an open question what kind of dynamic phenomena may occur during slip motion and how they can be  predicted. The only exception is the dynamics of one single point contact in two dimensions, which has been analyzed in detail by several authors. Such systems may undergo a singularity called dynamic jam during slip \citep{Genot1999}, in addition to the classical slip-stick, slip-reversal and slip-liftoff transitions. The same systems may also exhibit other forms of contact dynamics such as self-excited bouncing motion (a.k.a inverse chattering) \citep{paper2} and impulsive contact forces or "impact without collision" \citep{LeSuanAn_book,Zhao15} albeit transition into such modes does not occur during slip motion.

The goal of the present paper is to review these results and to investigate systematically the generic transitions and singularities of  multicontact systems. We perform a detailed analysis of systems with two slipping point contacts in two dimensional physical space, and many new phenomena are discovered. Our new findings include
\begin{itemize}
\item the sudden onset of impulsive contact forces (i.e. impact without collision or IWC) blocking slip motion
\item the onset of self-excited oscillations during slip, which may either remain microscopic, or they may grow exponentially to a macroscopic scale
\item convergence to two different types of codimension 2 singular manifolds where the dynamics becomes ill-defined within rigid body theory. One of these singularities is analogous to dynamic jamming, whereas the other one does not have an analogue in the single-contact case.
\end{itemize}
We then turn to systems with arbitrary number of point contacts and highlight some general properties of the dynamic jamming singularity.

There are two classical ways to deal with the piecewise smooth character of rigid unilateral contacts and Coulomb friction. The appealing mathematical framework of complementarity problems and measure differential inclusions offers a unified approach \citep{Leine_book,Brogliato1999} or one can  distinguish between various contact modes (slip, stick and separation) each of which implies different smooth behavior. Since our goal includes predicting and understanding transitions between various contact modes, it is a natural choice here to follow the contact mode-based approach. 

It is well-known that rigid body theory cannot be developed into a complete and consistent modeling framework in the presence of contacts. Solution inconsistency and indeterminacy have been known for a long time \citep{Jellet1872,Painleve1895}, and many examples of this phenomenon have been found and studied more recently \citep{champneys2016painleve}. In addition,  the solution may become ill-defined at attractive singular points \citep{Genot1999,szalai2014nondeterministic}. The limitations of rigid models can often be resolved by contact regularization: the rigid contacts are replaced by compliant ones, and the behavior of the regularized system is studied in the quasi-rigid limit $\epsilon\rightarrow 0$ of a compliance parameter $\epsilon$. We will make extensive use of contact regularization in the present paper.

In Sec. 2, we introduce our notation (Sec. 2.1) and review relevant results of the contact mode-based approach and of contact regularization wherein we follow the previous works \citet{champneys2016painleve} and \citet{varkonyi2017dynamics}. We also introduce a conceptual model system in Sec. 2.4 (inspired by a similar model system from \citet{nordmark2017dynamics}), which is used to illustrate certain phenomena by numerical simulation as well as to construct examples of other phenomena analytically. 

Sec. 3 focuses on the transitions of systems with slipping contacts. Existing results regarding the case of $n=1$ point contact are reviewed first, and we also formulate a theorem, which shows how and to what extent these systems stay away from those states where rigid models become ill-defined. We then turn towards systems with $n=2$ point contacts where a detailed analysis of transitions is given. Finally, we  discuss the case of general $n$, and a theorem related to dynamic jamming is generalized to this case. The paper is closed by a Discussion section, and by an appendix where some technical proofs are presented.

\section{Instantaneous behaviour of systems with sliding contacts}

\subsection{General formulation}
We consider the motion of an autonomous  mechanical system consisting of one or more rigid elements in two dimensions, subject to any number of ideal constraints (i.e. bilateral, frictionless connections) as well as $n\geq 1$ unilateral point contacts. We assume Coulomb friction with given coefficients of friction at all contact points. We use lowercase letters for scalars and vectors, whereas capital letters denote matrix quantities. The state of the system is characterized by a vector of generalized coordinates $q$ as well as the generalized velocities $\dot q$.
The continuous dynamics of such a system is given by a differential equation of the form
\begin{align}
\ddot{q}=f(q,\dot q)+G_T(q)\lambda_T+G_N(q)\lambda
\label{eq:basiceq}
\end{align}
where $\lambda\in\mathbb{R}^n$ is a column vector of $n$ non-negative normal contact forces $\lambda_i$ and $\lambda_T$ is a similar vector containing tangential contact forces. The contact forces are determined by unilateral contact constraints and by the friction model. (Note that the equation above is not valid during impacts, since then the contact forces are impulsive.)

Let $x_i(q)$ and $z_i(q)$ denote tangential and normal coordinates of the point contacts such that $z_i=0$ corresponds to a closed contact and $z_i>0$ to an open one, and $z_i=\dot z_i=\dot x_i=0$ means stick. 

We will examine slip motion, which means that contacts are initially closed with nonzero tangential velocity. We assume without loss of generality that $\dot x_i>0$ for all $i$. Then, the tangential contact forces $\lambda_{t,i}$ become dependent on the normal forces $\lambda_{i}$ by Coulomb's law
$$
\lambda_{t,i}=-\mu_i(q)\lambda_i
$$
(where $\mu_i$  is known and velocity-independent). In this case, \eqref{eq:basiceq} can be written as
\begin{align}
\ddot{q}=f(q,\dot q)+G(q)\lambda
\label{eq:basiceq2}
\end{align}

If the functions $x_i(q)$, $z_i(q)$, $f(q,\dot q)$, and $G(q)$ are known, then one determine two sets of equations of the form
\begin{align}
\ddot x&=a(q,\dot q)+K(q)\lambda
\label{eq:xdyn}\\
\ddot z&=b(q,\dot q)+P(q)\lambda
\label{eq:zdyn}
\end{align}
describing the dynamics of the contact points.
Here, $z=[z_1,z_2,...,z_n]^T$, $x=[x_1,x_2,...,x_n]^T$. The vectors $a,b\in\mathbb{R}^n$ and the $n$-by-$n$ matrices $K,P$ depend on the system in question. $P$ is sometimes referred to as the Delassus matrix of the system. 

\begin{figure}[h]
\begin{center}
\includegraphics [width=4cm] {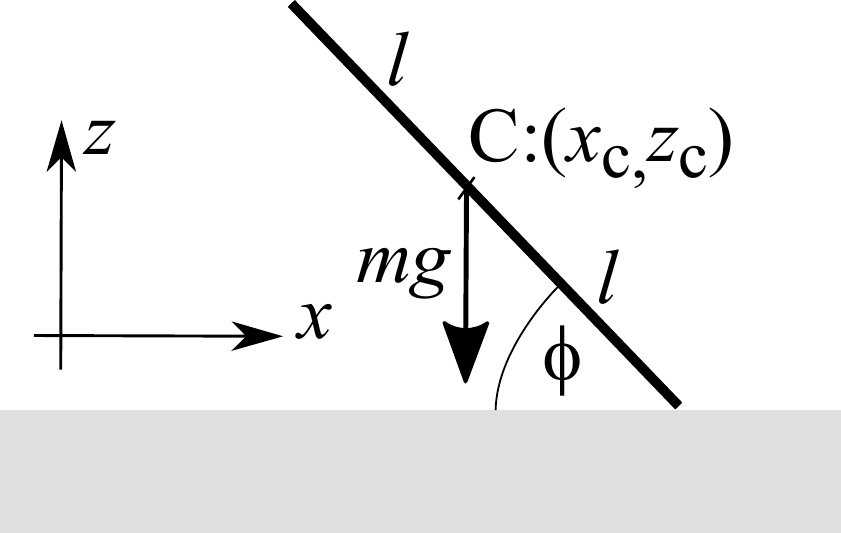}
\caption{Rod slipping at one endpoint}\label{fig:rod}
\end{center}
\end{figure}

\textbf{Example:} in the case of a  rigid rod (Fig. \ref{fig:rod}) of length $2l$, mass $m$, radius of inertia $\rho$ under gravitational force $mg$, with one endpoint in unilateral contact with a flat, horizontal surface, we can use the generalized coordinates $q=[x_c,z_c,\phi]^T$. The equations of motion are
\begin{align}
\ddot x_c&=-m^{-1}\mu\lambda\\
\ddot z_c&=-g+m^{-1}\lambda\\
\ddot\phi&=m^{-1}\rho^{-2}l\lambda(-\cos\phi+\mu\sin\phi)
\end{align}
and the contact coordinates can be expressed as
\begin{align}
x&=x_c+l\cos\phi\\
z&=z_c-l\sin\phi
\end{align}
 The Lagrange equations and the kinematics of the rod then imply \citep{Genot1999}
\begin{align}
a&=-l\dot\phi^2\cos\phi\\
b&=l(\dot\phi^2\sin\phi-g)\label{eq:rodb}\\
K&=m^{-1}\left[-1+l^2\rho^{-2}(-\mu\sin^2\phi+\cos\phi\sin\phi)\right]\\\\
P&=m^{-1}\left[1+l^2\rho^{-2}\cos^2\phi-\mu\cos\phi\sin\phi\right]\label{eq:rodP}\\
\end{align}
where $\mu$ is the coefficient of friction . Note that $a,b,K,P$ are all scalars, for now $n=1$.

In what follows, we denote the $j^{th}$ column vector of $K,P$ by $k_j$ and $p_j$ and the $i^{th}$ element of vectors $a,b,k_j,p_j$ by $a_i,b_i,k_{ij},p_{ij}$.
We drop arguments like $t,q(t),\dot q(t)$ for brevity. We will use a lower index $_{init}$ to denote initial conditions at some initial time $t_{init}<0$ and index $_{0}$ for quantities evaluated at $t=0$. Typically, we will choose initial conditions in such a way that transitions from slip occur at $t=0$. 

Slip motion requires an initial state $q_{init},\dot q_{init}$ inducing $z_{init}=\dot z_{init}=0$, $\dot x_{init}>0$ (where the inequality should be satisfied by all elements of vector $x$). In this situation, each contact point may undergo sustained slip in the positive $x_i$ direction (contact mode S) or lift-off (contact mode F). Stick and negative slip  are not possible since $\dot x_{init}>0$. Each one of these contact modes has a corresponding equality and an inequality constraint:
\begin{align}
\ddot z_i=0
\label{eq:slip-equality}\\
\lambda_i\geq 0
\label{eq:slip-inequality}
\end{align}
 for S mode and 
 \begin{align}
\lambda_i=0
\label{eq:liftoff-equality}\\
\ddot z_i\geq 0
\label{eq:liftoff-inequality}
\end{align}
for F mode. The contact mode of a full system can be represented by an $n$-letter word from the two-letter alphabet $\{S,F\}$. We can check the feasibility of all $2^n$  contact modes by solving \eqref{eq:zdyn} together with the $n$ equality constraints, and by testing the inequality constraints of the contact modes. The fact that we may find multiple or no feasible contact mode is known as Painlev\'e paradox \citep{champneys2016painleve}. This kind of limitation of rigid models is the primary motivation for contact regularization techniques introduced below.

\subsection{Contact regularization}

The non-uniqueness and non-existence of solution associated with rigid models is often resolvable by contact regularization techniques. A regularized contact model allows small penetrations $z_i<0$ at the contact points and assumes a certain relation between the contact force and the penetration. Here, we will use a standard, linear, unilateral viscoelastic contact law 
\begin{align}
\lambda_i=
\begin{cases}
0 & if\;\;  z_i>0\\
\max(0,-\epsilon^{-2}k_iz_i-\epsilon^{-1}\nu_i\dot z_i) & if\;\;  z_i \leq 0
\end{cases}
\label{eq:contactlaw}
\end{align}
where, $k_i$ and $\nu_i$ are scaled stiffness and damping coefficients, whereas $\epsilon$ is a scaling factor. Now we will say that a contact is closed if $\lambda_i>0$ (as opposed to $z_i=0$ in the rigid case). Note that the contact model is smooth as long as the contact is closed.

Then, the analysis of a rigid model is replaced by the quasi-rigid limit $\epsilon\rightarrow 0$ of the regularized model, which often yields deeper insight into the behavior of the system. The regularized model is a slow-fast system in which $a,b,P,K$ evolve on a slow time scale, and the internal dynamics of contacts ($z,\dot z$) represents the fast subsystem. In Section 2 of the paper, we will focus exclusively on the fast dynamics and we do not even specify  the full slow-fast dynamics or the reduced problem. Then in Section 3, we discuss sudden transitions of the fast dynamics induced by the slow dynamics.

When all $n$ contact points are closed ($\lambda_i>0$ for $i=1,2,...,n$), then  the equations \eqref{eq:zdyn} and \eqref{eq:contactlaw} induce the fast dynamics:
\begin{align}
g' = 
\begin{bmatrix}
O_n & I_n \\
 -PK & -PN  
\end{bmatrix}
g+
\begin{bmatrix}
o_n,\\b  
\end{bmatrix}
\label{eq:contactdynamics_generaln}
\end{align}
where $'$ represents differentiation with respect to fast time $\tau=\epsilon^{-1}t$; $O_n$ and $I_n$ are zero and identity matrices of size $n\times n$, $o_n$ is a column vector of $n$ zeros, and
\begin{align}
g=
\epsilon^{-2}\begin{bmatrix}
z\\
z'  
\end{bmatrix},
K=
\begin{bmatrix}
k_1&0&...&&0\\
0&k_2&0&...&0\\
\vdots&&&&\\
0&...&&0&k_n 
\end{bmatrix},
N=
\begin{bmatrix}
\nu_1&0&...&&0\\
0&\nu_2&0&...&0\\
\vdots&&&&\\
0&...&&0&\nu_n 
\end{bmatrix}
\end{align}
Note that $g$ is a rescaled vector depending on those variables, which represent motion in the directions of the contact normals. Such a reduction is possible because the fast dynamics of $g$ decouples from all other components of the dynamics including the dynamics of $x$ and its time derivative. 

When the regularized model is used, we identify slip motion with motion during which all contacts are closed and the fast dynamics remains stationary at the invariant point 
$$
g=-\begin{bmatrix}
O_n & I_n \\
 -PK & -PN  
\end{bmatrix}^{-1}
\begin{bmatrix}
o_n,\\b  
\end{bmatrix}
$$
of the linear system \eqref{eq:contactdynamics_generaln}. Such an invariant point may be asymptotically stable or unstable depending on the eigenvalues of the coefficient matrix. Hence, regularization shows that certain modes involving slipping contacts are unstable and do not occur in practice, which can eliminate some forms of solution non-uniqueness. 


As we point out later, contact regularization has other benefits as well. Most importantly, we will encounter situations, when the invariant point is asymptotically unstable and  $z_i$ diverges towards $-\infty$. Such an event induces rapidly increasing contact force $\lambda_i$. We will identify this kind of behaviour with the impact without collision (IWC) phenomenon of rigid models. Detailed analysis of dynamics induced by diverging contact forces (similar to \citet{Zhao15} in the case of $n=1$) is however beyond the scope of this work.

\subsection{Systems with a single point contact}\label{sec:1contact-instantaneous}
In the case of $n=1$\citep{champneys2016painleve}, $a,b,P,K,\lambda$ are scalars. Slip motion means that 
\begin{align}
\lambda=-b/P
\label{eq:lambda-1contact}
\end{align}
according to \eqref{eq:zdyn} and \eqref{eq:slip-equality}. Then the feasibility condition \eqref{eq:slip-inequality} is satisfied if $b$ and $P$ have opposite signs. Similarly, liftoff is feasible if $b>0$. Since both $b$ and $P$ may have any sign, either one, both or none of the two modes may be feasible (Table \ref{table:1contact}). The issue of non-existence is resolved by considering a third type of solution involving impulsive contact forces. Whenever $P<0$, an impact without collision (IWC) may occur: an impact occurs, which causes instantaneous jump in tangential velocity into stick or reverse slip \citep{LeSuanAn_book,Zhao15,hogan2017regularization}. The regularized contact model \eqref{eq:contactlaw} predicts rapid divergence of $\lambda$ to $\infty$ in such situations, which is consistent with the assumption of an IWC in the $\epsilon\rightarrow 0$ limit.

Stability analysis using regularized contact model reveals that slip is stable if and only if $P>0$, yet stability analysis is not able to eliminate non-uniqueness if $P<0<b$.\\

\textbf{Example:} the parameters $b,P$ of the sliding rod (Fig.\ref{fig:rod}) given by \eqref{eq:rodb},\eqref{eq:rodP} may have either sign. In particular, $b$ is always negative if $\dot\phi$ is small, but may become positive for large $\dot\phi$, whereas $P$ is always positive for small $\mu$ and may have either sign if $\mu$ is large. 

\begin{table} 
\centering
\caption{Feasibility, and stability of contact modes and IWC in the case of one point contact}
\label{table:1contact}
\begin{tabulary}{0.6\textwidth}{ccc}
\hline
& b\textless 0 & b\textgreater 0                                                            \\ 
\hline
P\textless 0    & IWC         & \begin{tabular}{c}IWC\\ liftoff\\ slip (unstable)\end{tabular}   
\\ \hline
P\textgreater0 & liftoff     & slip (stable)                                                             \\ \hline
\end{tabulary}
\end{table}

\subsection{Systems with two point contacts}\label{sec:2contact-instantaneous}

\begin{figure}[h]
\begin{center}
\includegraphics [width=6cm] {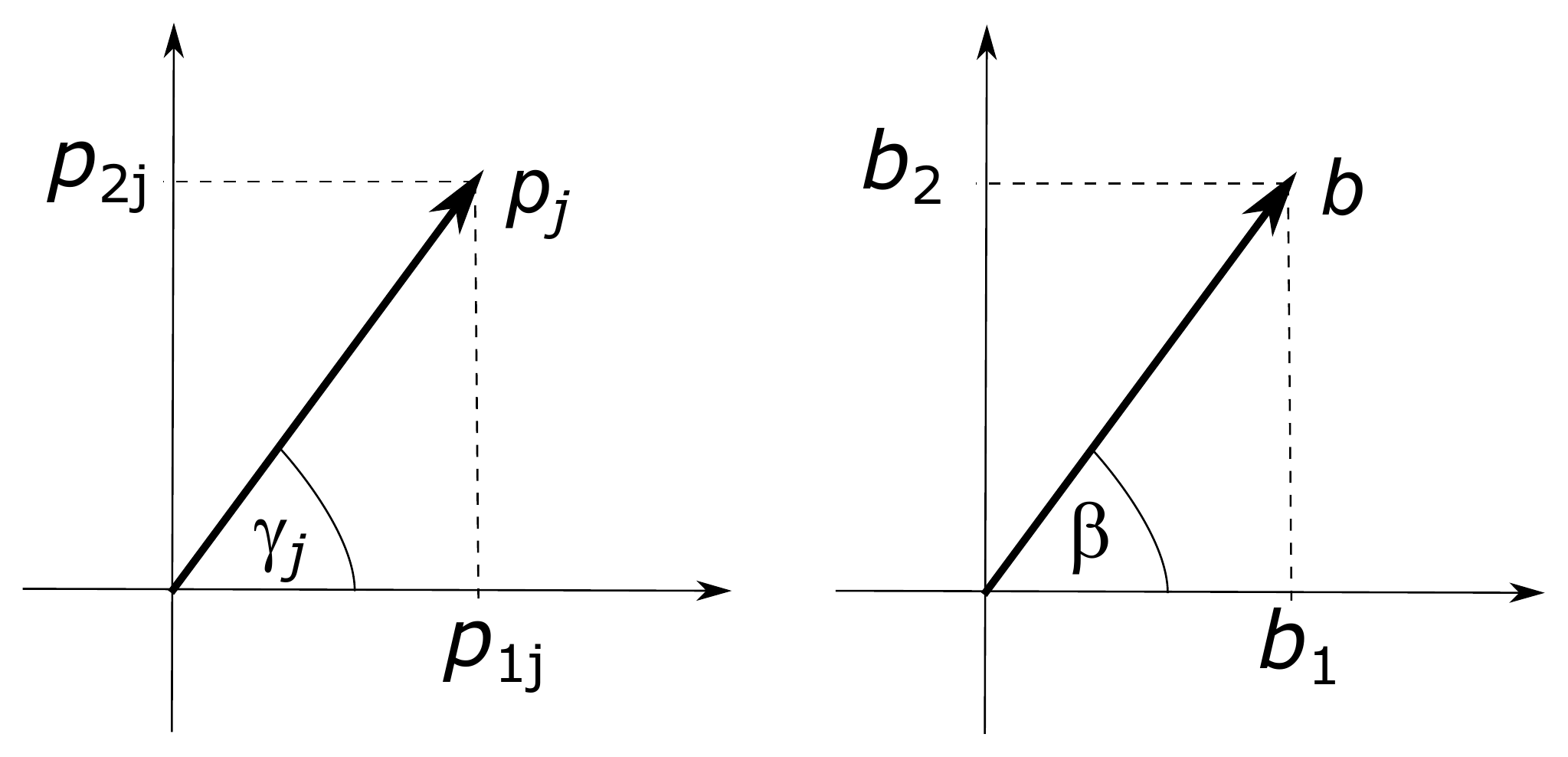}
\caption{Definition of the angles $\gamma_i$ ($i=1,2$) and $\beta$.}\label{fig:gammabeta}
\end{center}
\end{figure}

\begin{figure}[h]
\begin{center}
\includegraphics [width=13cm] {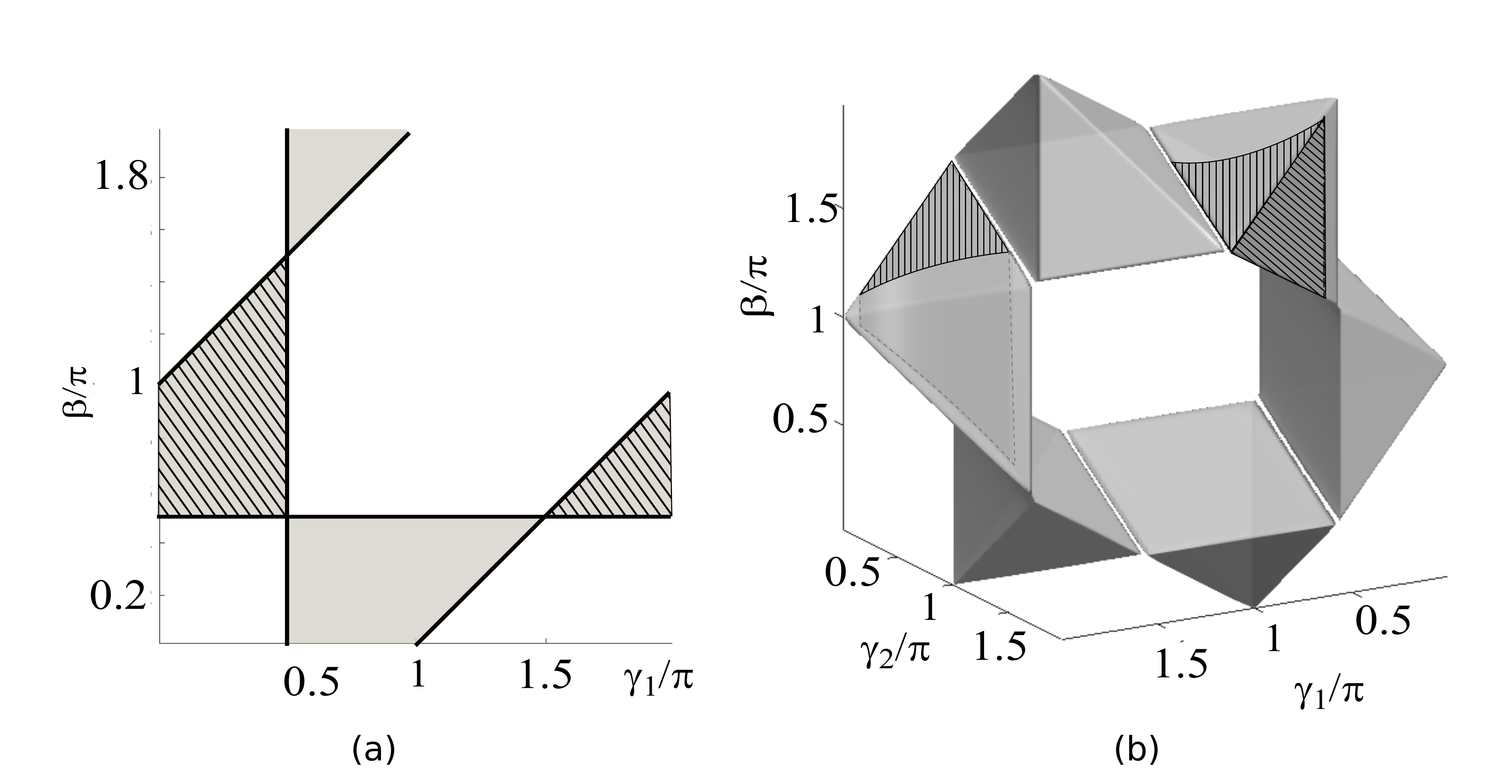}
\caption{Feasibility and stability of the contact modes SF (a) and SS (b). Grey areas or volumes show feasibility and hatching marks regions where motion is also stable. } \label{fig: 2contact feasibility}
\end{center}
\end{figure}

In the case of 2 contacts, there are 4 possible contact modes: SS, FF, SF, and FS. The feasibility and stability analysis of these systems was recently investigated by  \citet{varkonyi2017dynamics}. The analysis reveals that the most important model parameters are the angle $\beta$ between vector $b$ and the unit vector $[1,0]$ (measured in the direction of $[0,1]$), as well as the analogous angles $\gamma_1$ and $\gamma_2$ associated with the vectors $p_1$, $p_2$ (Figure \ref{fig:gammabeta}). The feasibility of the four contact modes depends exclusively on these three parameters. For example, FF motion (liftoff at both points) is feasible if $b_1,b_2>0$, i.e. if both points accelerate in the absence of contact forces away from the contact surfaces. This can also be expressed as $0\leq\beta\leq\pi/2$. Contact mode FS is feasible if we have positive contact force at point 2, and positive normal acceleration at point 1 i.e. if
\begin{align}
-b_2/p_{22}>0 \label{eq:FS1}\\
b_1-p_{21}b_2/p_{22}>0 \label{eq:FS2}
\end{align}
As we know from Sec. \ref{sec:1contact-instantaneous}, the stability of FS additionally requires $p_{22}>0$. These conditions depend exclusively on the angles $\gamma_1,\beta$ (Fig. \ref{fig: 2contact feasibility}(a)). The feasibility and stability conditions of SF are analogous.

For simultaneous slip at both contact points, the contact force is determined by \eqref{eq:zdyn}:
\begin{align}
\lambda=-P^{-1}b
\label{eq:lambda-2contact}
\end{align}
The feasibility condition \eqref{eq:slip-inequality} is satisfied exactly when $-b$ is in the cone spanned by $p_1$ and $p_2$, which can be expressed in terms of the angles $\beta,\gamma_1,\gamma_2$ as illustrated by Fig. \ref{fig: 2contact feasibility}(b). The figure also shows the region of stability obtained by eigenvalue analysis. This region will soon be analyzed in more detail (see Fig. \ref{fig:SSrange} below) 

The feasibility of impulsive contact forces (IWC) has also been examined in \citet{varkonyi2017dynamics}. It was found that there are three different types of IWC since point 1, point 2 or both points may exhibit impulsive contact forces. Unlike in the case of $n=1$, contact regularization does not yield uniform answer with respect to the feasibility of IWCs: in certain cases, the feasibility of the IWC may depend on the stiffness and dissipation coefficients $k_1,k_2,\nu_1,\nu_2$ of the regularized contact model.

The stability analysis of the contact modes  via contact regularization yields the following results:
\begin{enumerate}
\item the stability of the SF and FS modes is  determined by $\gamma_1,\gamma_2$ and it is independent of any other parameter. The same is true for SS in a large range of $\gamma_1,\gamma_2$
\item in certain ranges of $\gamma_1,\gamma_2$, the stability of the SS mode may also depend on the model parameters $k_1,k_2,\nu_1,\nu_2$
\item stability of any mode is independent of $\beta$

\end{enumerate}

Finally, \citep{varkonyi2017dynamics} shows that the regularized model predicts unusual forms of contact dynamics in some ranges of $\gamma_1,\gamma_2,\beta$, including limit cycles of the fast dynamics, and self-excited exponentially growing oscillations. A comprehensive list of such dynamic phenomena is not known.

\subsection{An illustrating example}\label{sec:numericexample}
We will illustrate several types of dynamic behavior by analytical investigation and numerical simulation of a conceptual model system with $n=2$ contacts. Our model captures the algebraic structure of the equations of motion, but does not correspond to a specific mechanical system. 

We have seen that $P$ is a function of $q$ whereas $b$ may additionally depend on $\dot q$. Similarly to the equations \eqref{eq:xdyn}-\eqref{eq:zdyn} developed for the dependent variables $x$ and $z$, we can also develop a set of expressions for $\dot P$ and $\dot b$ by using \eqref{eq:basiceq2} and the chain rule. In the present paper, we do not explicitly construct such equations for any system, but we note that they can always be written as
\begin{align}
\dot P = A_1(q,\dot q)\\
\dot b =  \alpha_2(q,\dot q)+ A_3(q,\dot q)\lambda 
\end{align}
where the functions $A_1,A_3\in\mathbb {R}^{2\times 2}$, $\alpha_2\in\mathbb {R}^{2}$ are system-specific. For example, the  values of $A_1,\alpha_2,A_3$ corresponding to a frictional impact oscillator originally introduced by \citet{leine2002} are shown in Appendix B of \citet{nordmark2017dynamics}.

For the sake of illustration, we will now consider the case when $A_1,\alpha_2,A_3$ are constants:
\begin{align}
\dot P = \alpha_1\label{eq:dotP}\\
\dot b =  \alpha_2+ A_3\lambda \label{eq:dotb}
\end{align}
This model system can be combined with the rigid contact model, where we have $\lambda=0$ in F mode and $\lambda$ is given by \eqref{eq:lambda-2contact} in S mode. Then the equations \eqref{eq:dotP}-\eqref{eq:dotb} decouple from all other equations of motion, thus we are able to investigate the dynamics of $P$ and $b$ in isolation. This system will be used to demonstrate the onset of some phenomena analytically. 
Alternatively, we can also combine \eqref{eq:dotP}-\eqref{eq:dotb} with the regularized contact model. Then, the equations \eqref{eq:zdyn}, \eqref{eq:contactlaw}, \eqref{eq:dotP} and\eqref{eq:dotb} decouple from all other equations of motion. We will use this system to illustrate our findings by a series of numerical simulations (Fig. \ref{fig:sim1}-Fig. \ref{fig:sim4}). In each figure, we show the time history of $z_1$ and $z_2$ with circles denoting liftoff and landing at one of the contact points. In some of the figures, the time history of the angles $\beta$, $\gamma_i$ is also shown. The model parameters and initial conditions used in these simulations are given in the appendix.   

\section{Generic transitions from slip}
The main goal of the paper is to identify and to give a qualitative description of generic transitions and singularities of a system, which is initially in slip mode. Clearly, such events occur when the actual contact mode of the system becomes either unfeasible or unstable. In what follows, we begin with the $n=1$ case, where existing results are reviewed and some new results are also presented. This is followed by detailed analysis for $n=2$ and some results for general $n$.

\subsection{Transitions of systems with a single point contact}

Assume now that a system with a single point contact undergoes slip motion with nonzero velocity
 and this mode of motion is feasible and stable. As we know from Sec. \ref{sec:1contact-instantaneous}, this means 
\begin{align}
\dot x_1&>0
\label{eq:1pointslidingcondition1}\\
b&<0<P
\label{eq:1pointslidingcondition2}
\end{align} 
Since $\dot x_1,b$ are functions of all state variables ($q$ and $\dot q$) and $P$ depends on $q$, all three variables vary over time.

A sudden transition occurs, if one or several of the conditions \eqref{eq:1pointslidingcondition1}-\eqref{eq:1pointslidingcondition2} break down. There are two trivial transitions. First, reaching the border $\dot x_1=0$ means transition to stick or slip reversal. Moreover,\citet{nordmark2017dynamics} show using regularized contacts that this transition is in some cases immediately followed by exponentially growing bouncing motion of $z$, which is termed by those authors \textit{reverse chatter}. 
The second simple option is reaching the \textit{liftoff manifold} $b=0$, where slip becomes infeasible and the contact lifts of. It is tempting to declare that crossing $P=0$ (where slip again becomes infeasible) is also a generic event. Nevertheless $P=0$, which we will refer to the  \textit{\Pain manifold}, corresponds to a singularity of the contact force according to \eqref{eq:lambda-1contact}. What happens instead was uncovered first by \citet{Genot1999}. They studied the problem of a sliding rod, and found that as $P$ approaches 0, the large contact force induces large acceleration of the rod. The value of $b$ is a function of the velocity $\dot q$ whereas $P$ is independent of it. In the case of the rod, $b$ rapidly increases in response to large contact forces and thus it may approach $b=0$. Thereby the system may converge to the codimension-2 manifold given by $b=P=0$. We will refer to this as the \textit{G\'enot-Brogliato manifold}. Here, the contact force is left ill-defined by\eqref{eq:lambda-1contact}. Local analysis of the system near the \GB manifold reveals that it attracts an open set of initial conditions, and the contact force diverges to infinity, which inspired the name \textit{dynamic jamming} of this phenomenon. What happens beyond that point is a subtle question, which was only partially uncovered by using asymptotic analysis and geometric singular perturbation techniques \citep{nordmark2017dynamics,kristiansen2017canard}.

The local analysis of the rod was extended in unpublished work by Nordmark, Dankowicz, and Champneys (see also Fig. 13 in \citet{champneys2016painleve}) to general systems with a single point contact. They found that the G\'enot-Brogliato manifold is often attractive and the contact force may or may not diverge to infinity when approaching it. 

These local results suggest that the $P=0$ boundary is never crossed by  systems with $n=1$ away from the \GB manifold. Nevertheless a global extension of this result has never been published. Here we give a formal proof of this property:

\begin{theorem}
Assume that a mechanical system  with a single point contact is initially in slip mode, in a state with $b<0<P$, furthermore $P$ is Lipschitz with respect to $q$ and $K$, $a$, $b$ are continuous functions of their arguments. Then slip motion of the system may not reach a state where $ P = 0 <  b$
\label{thm: no_painleve_1_contact}
\end{theorem}

The proof of this statement is based on the observation that crossing the boundary would induce a $1/x$ type singularity in the contact force, which is non-integrable. This implies that the contact force would eliminate slip motion for any finite initial tangential velocity. The detailed proof is given below.
\begin{proof} [Proof by contradiction:] 

Assume that the system is launched at $t_{init}<0$ and a state with $P=0> b$ is reached at time $t=0$.

The velocity $\dot q$ is a continuous function of time during impact-free motion, hence $|\dot q|$ has a global maximum $\dot q_{max}$ over the closed interval $t\in[t_{init},0]$. This means that the following bound applies to $P$:
\begin{align}
0<P(q(t))<\mathcal{L}|q(t)-q_{0}| \leq \mathcal{L}\dot q_{max}(-t)
\label{eq:Pbound}
\end{align}
where $\mathcal L$ is a Lipschitz constant of $P$. This relation shows that the contact force \eqref{eq:lambda-1contact} undergoes an $f(x)=1/x$ type singularity.

Next we investigate how the tangential velocity of the contact point varies in response to the singular contact force. The tangential acceleration of the contact point can be expressed according to \eqref{eq:xdyn}, and \eqref{eq:lambda-1contact} as
\begin{align}
\ddot x&=a+K\lambda\\
&=a-KP^{-1}b
\end{align}
This can be rearranged as
\begin{align}
P\ddot x=Pa-Kb
\label{eq:tanaccformula}
\end{align}
implying
\begin{align}
(P\ddot x)|_{t=0}&=
\underbrace{P_0}_0a_0-K_{0}b_{0} <0
\end{align}
where the index $_0$ refers to values at $t=0$. The last inequality holds because $K_{0}<0$ follows from $P_{0}=0$, since any external force must cause an acceleration, whose scalar product with the force vector is positive (unless the force is directly opposed by a hard constraint). 

The right-hand side of \eqref{eq:tanaccformula} is continuous, thus there exists a time
$t_1$ sufficiently close to $0$ such that for all $t_1\leq t\leq 0$, $P\ddot x$ is strictly negative with some finite bound $E$:
\begin{align}
P\ddot x<-E<0
\end{align}

Then, we can combine this with \eqref{eq:Pbound} to obtain 
\begin{align}
\ddot x&<-P^{-1}E\\
     & < -\frac{E} {\mathcal{L}\dot q_{max}}(-t)^{-1}
\end{align}
This bound involves a non-integrable singularity, which means that the variation of $\dot x$ in the time interval $[t_1,0]$ is unbounded. In other words, for any positive sliding velocity at $t_1$, $\dot x = 0$ is reached for some $t<0$. Then a slip-stick transition occurs, contradicting the initial assumption that the system slips until $t=0$.
\end{proof}

We can summarize the results of this subsection as follows. There are four generic transitions from slip: transition to stick or slip reversal (possibly followed by inverse chattering); lift-off; and the dynamic jamming singularity. Crossing the \Pain manifold $P=0$ is impossible, i.e. such systems never evolve into states suffering from Painlev\'e paradox except for the codimension 2 \GB manifold. Notice that the last result required the Lipschitz property of $P$. This is violated if the friction coefficient jumps abruptly, in which case a system may slip into a state subject to \Pain paradox away from the \GB manifold.

\subsection{Transitions of  systems with two point contacts}
As before, we assume that the system under consideration is in slip state initially with nonzero tangential velocity at both contact points, furthermore slip motion is feasible and stable. In the next subsection, we find all generic boundaries at which feasibility and stability can be lost. Then the remainder of the section focuses on the question: what happens after reaching each of these boundaries?

\subsubsection{Feasibility and stability boundaries}
\begin{figure}[h]
\begin{center}
\includegraphics [width=8cm] {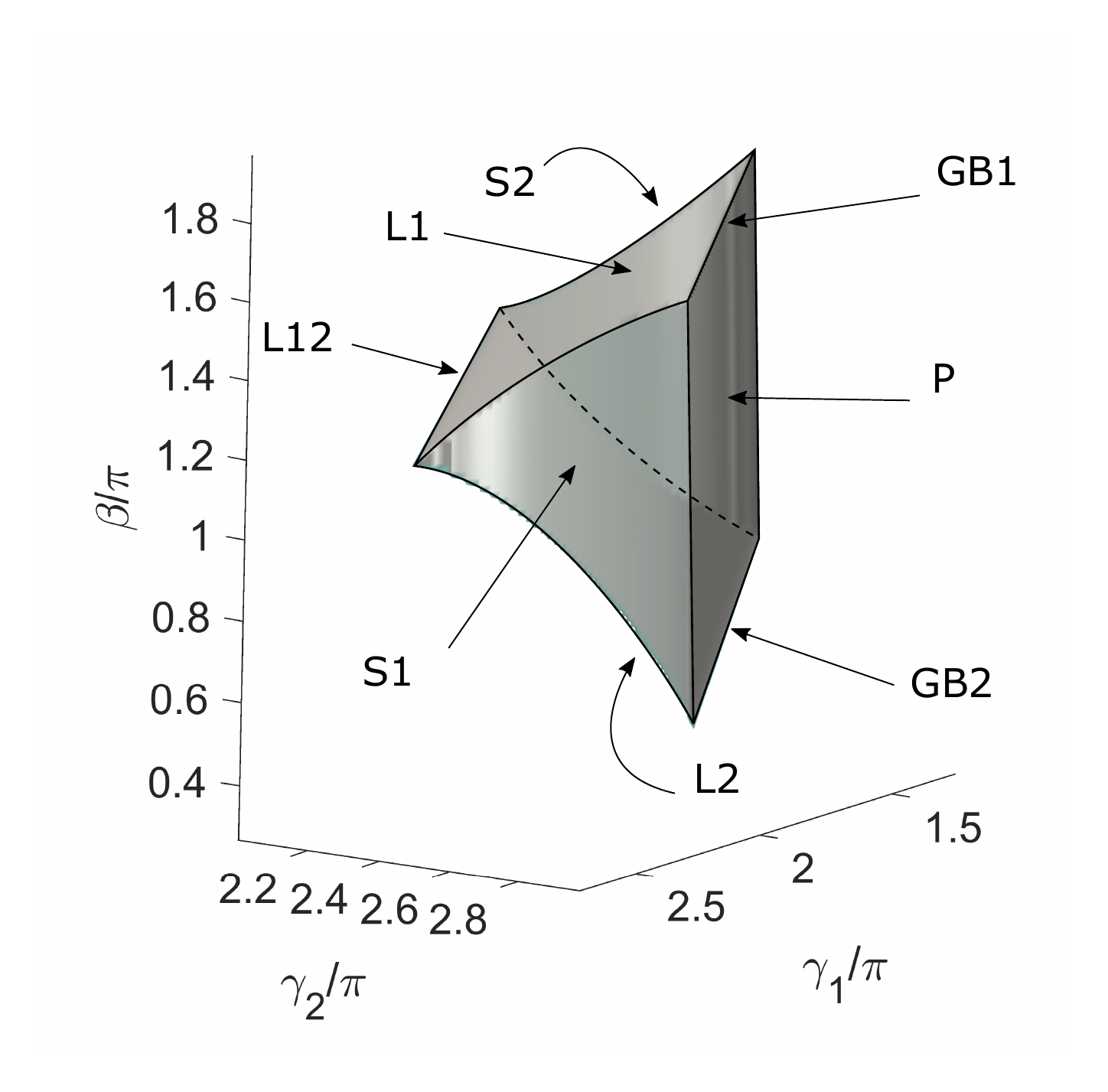}
\caption{The range of angles $\beta$,$\gamma_1$, $\gamma_2$ where the SS contact mode is feasible and stable. Note that the intersection of the coordinate axes is shifted from 0 for better visibility. The exact location of the curved surfaces S1, S2 depends on $|p_1|, |p_2|$, $k_1$, $k_2$, $\nu_1$, and $\nu_2$ The figure shows the case $|p_1|=|p_2|=k_1=1$, $k_2=10$, $\nu_1=\nu_2=0$. For more information about possible variations of the stability boundaries, see Fig. \ref{fig:L1parts}.}\label{fig:SSrange}
\end{center}
\end{figure}

\begin{figure}[h]
\begin{center}
\includegraphics [width=9cm] {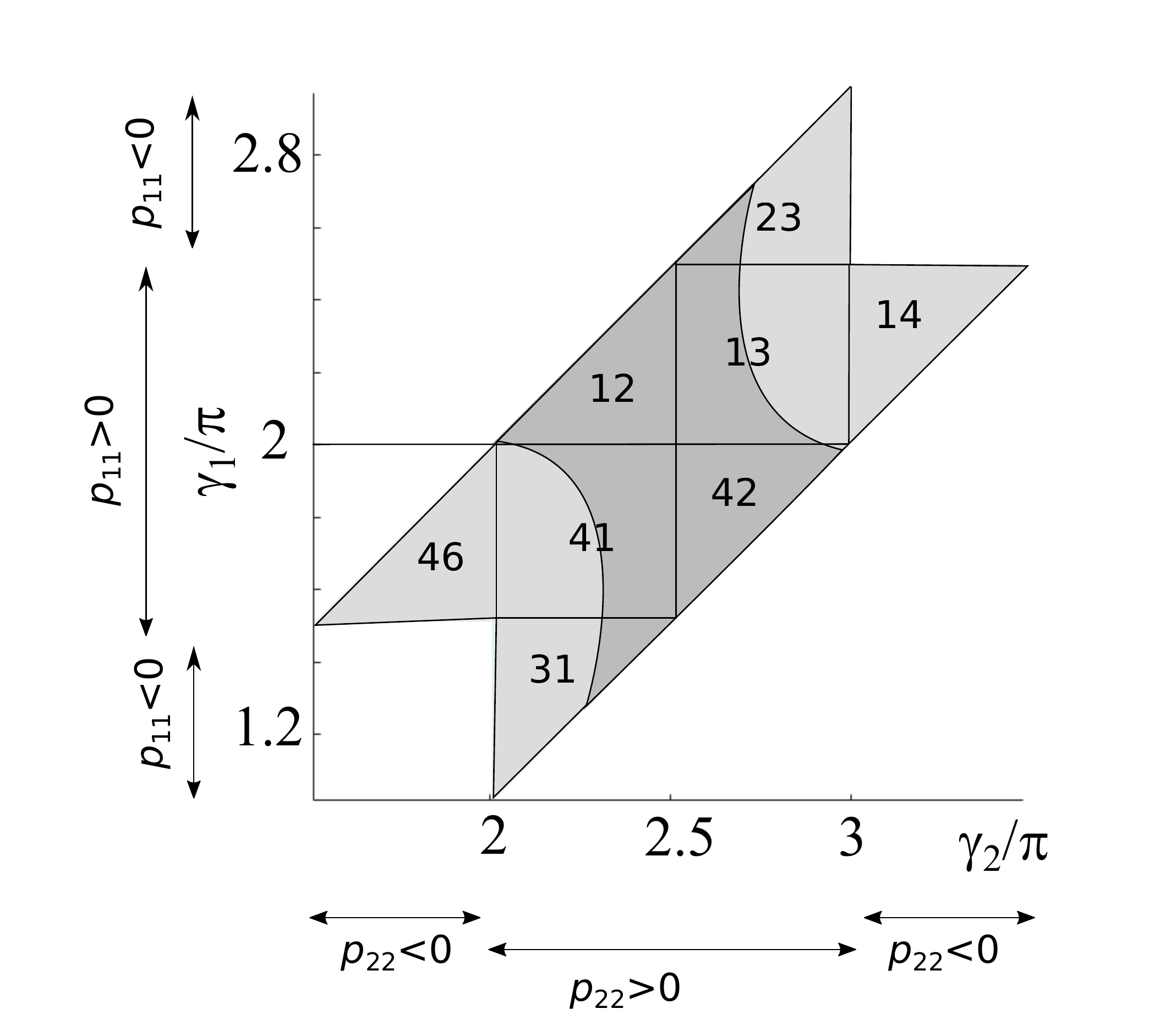}
\caption{Projection of Fig. \ref{fig:SSrange} to the $\gamma_1-\gamma_2$ plane (dark grey area), possible locations of the projection if parameters $|p_i|,k_i,\nu_i$ are varied (light grey area), and subdivision of the L1 surface (solid lines and two-digit numbers).}\label{fig:L1parts}
\end{center}
\end{figure}

Reaching zero tangential velocity is an obvious violation of feasibility. In addition to that, detailed analysis of \citet{varkonyi2017dynamics} shows that we have feasibility and stability if the system parameters $\beta$, $\gamma_1$, $\gamma_2$ are in  a certain range (Fig. \ref{fig:SSrange}). This range is bounded by 5 smooth surfaces as follows:
\begin{enumerate}
\item two planar \textit{liftoff boundaries} L1, L2 given by $\beta=\gamma_2$ (L1) and $\beta=\gamma_1$ (L2). One of the contact forces $\lambda_1$ and $\lambda_2$ becomes zero at each of these surfaces.
\item the \textit{Painlev\'e boundary} P given by $|\gamma_1-\gamma_2|=\pi$ where matrix $P$ is singular, and the (degenerate) cone spanned by $p_1$ and $p_2$ is a full line. Here, the contact forces \eqref{eq:lambda-1contact} become infinitely large.
\item the stability boundaries S1, S2, which are curved surfaces composed of straight lines parallel to the $\beta$ axis. The coefficient matrix of \eqref{eq:contactdynamics_generaln} has a pair of purely imaginary eigenvalues along these surfaces, and thus slip motion becomes unstable. The exact locations of these surfaces depend on parameters $|p_i|,k_i,\nu_i$ but they always lie within the boundaries shown in Fig. \ref{fig:L1parts}. 
\end{enumerate}
There are three additional curves along the boundary, which deserve special attention as they are associated with singular behaviour:
\begin{enumerate}
\item the line sections GB1,GB2 are at the intersections of L1 and L2 with P. As we will see, these are analogous to the G\'enot-Brogliato manifolds of the single-contact case.
\item the line L12 is at the intersection of the liftoff manifolds L1 with L2 and it is given by $\beta=\gamma_1=\gamma_2$. Along L12, $P$ is singular and the cone spanned by $p_1$ and $p_2$ is a half-line. 
\end{enumerate}

Slip motion persists until one of the feasibilility or stability conditions is violated. In what follows, we analyze the consequences of crossing each of the boundaries listed above.

\subsubsection{Reaching zero tangential velocity}
The system may undergo slip-stick transition or slip reversal, if the tangential velocity at one or both contact points drops to zero. Note that simultaneous zero crossing is generic in the presence of certain kinematic constraints, as for example in the case of a single rigid object with two point contacts. Similarly to the case of a single point contact \citep{paper2}, it is likely that the regularized contact model sometimes predicts self-excited bouncing motion resembling inverse chattering in this situation, however we do not examine this possibility in the present paper.

\subsubsection{Crossing L1} 
\begin{figure}[h]
\begin{center}
\subfigure[]{\includegraphics [width=0.32\textwidth] {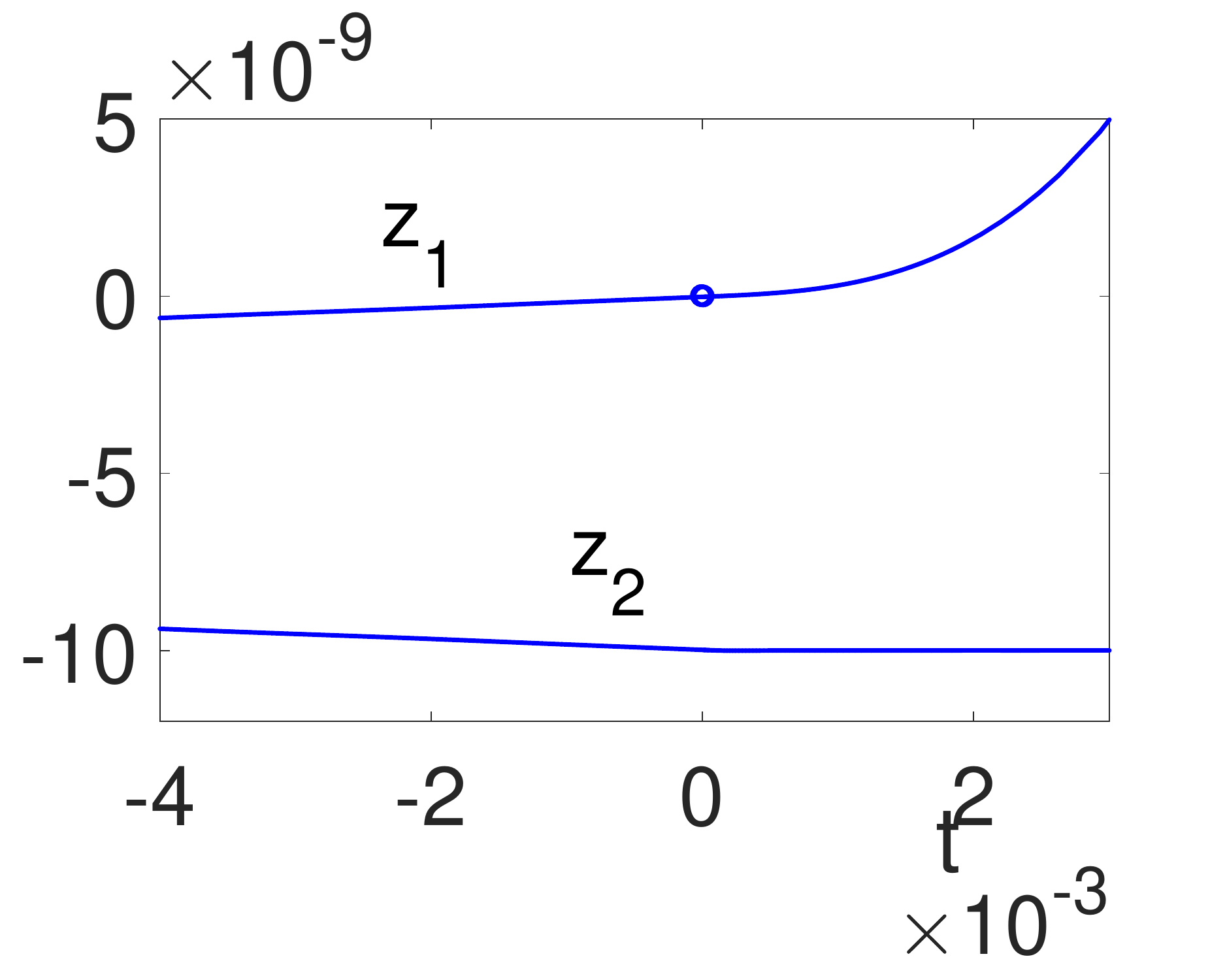}}
\subfigure[]{\includegraphics [width=0.32\textwidth] {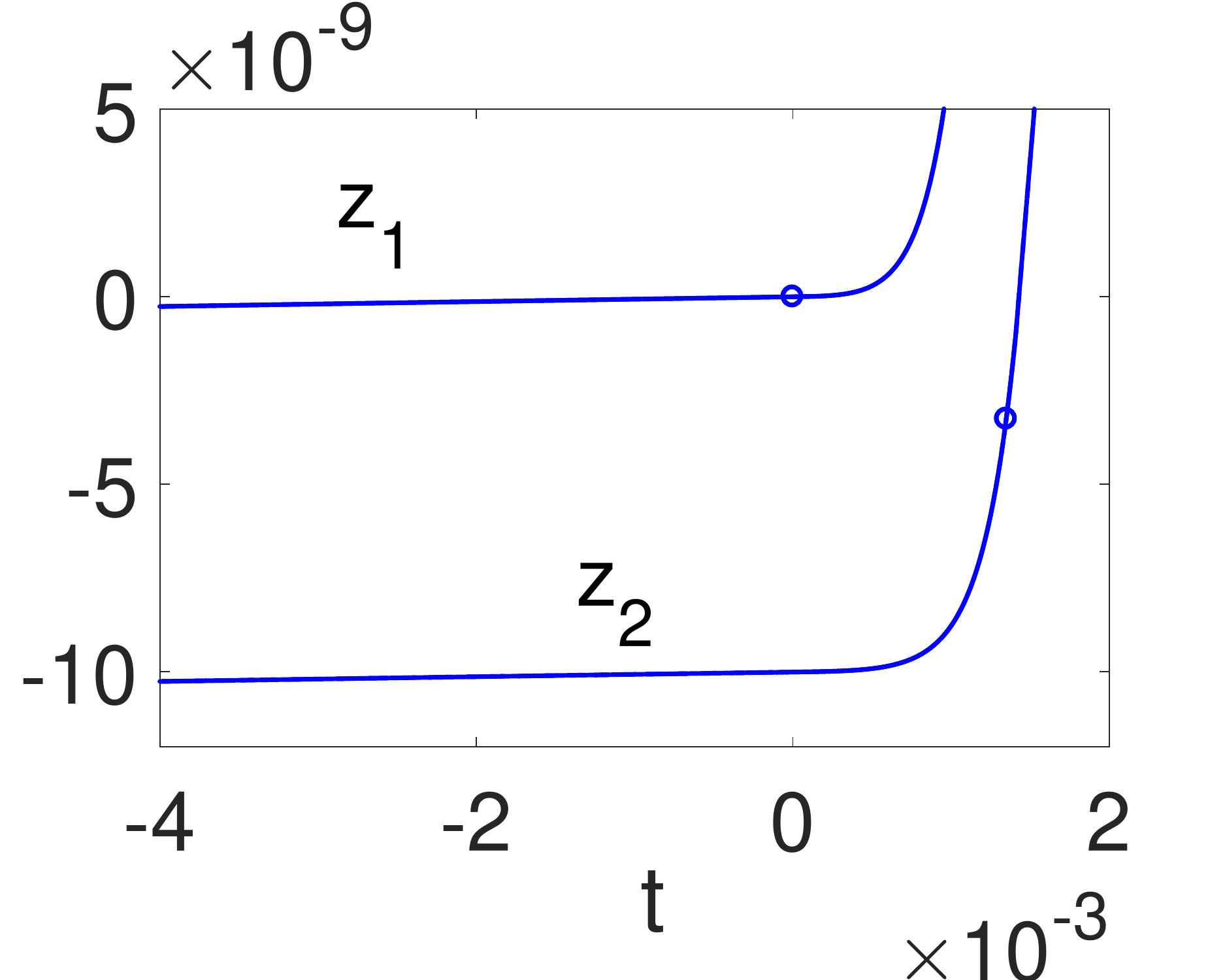}}
\subfigure[]{\includegraphics [width=0.32\textwidth] {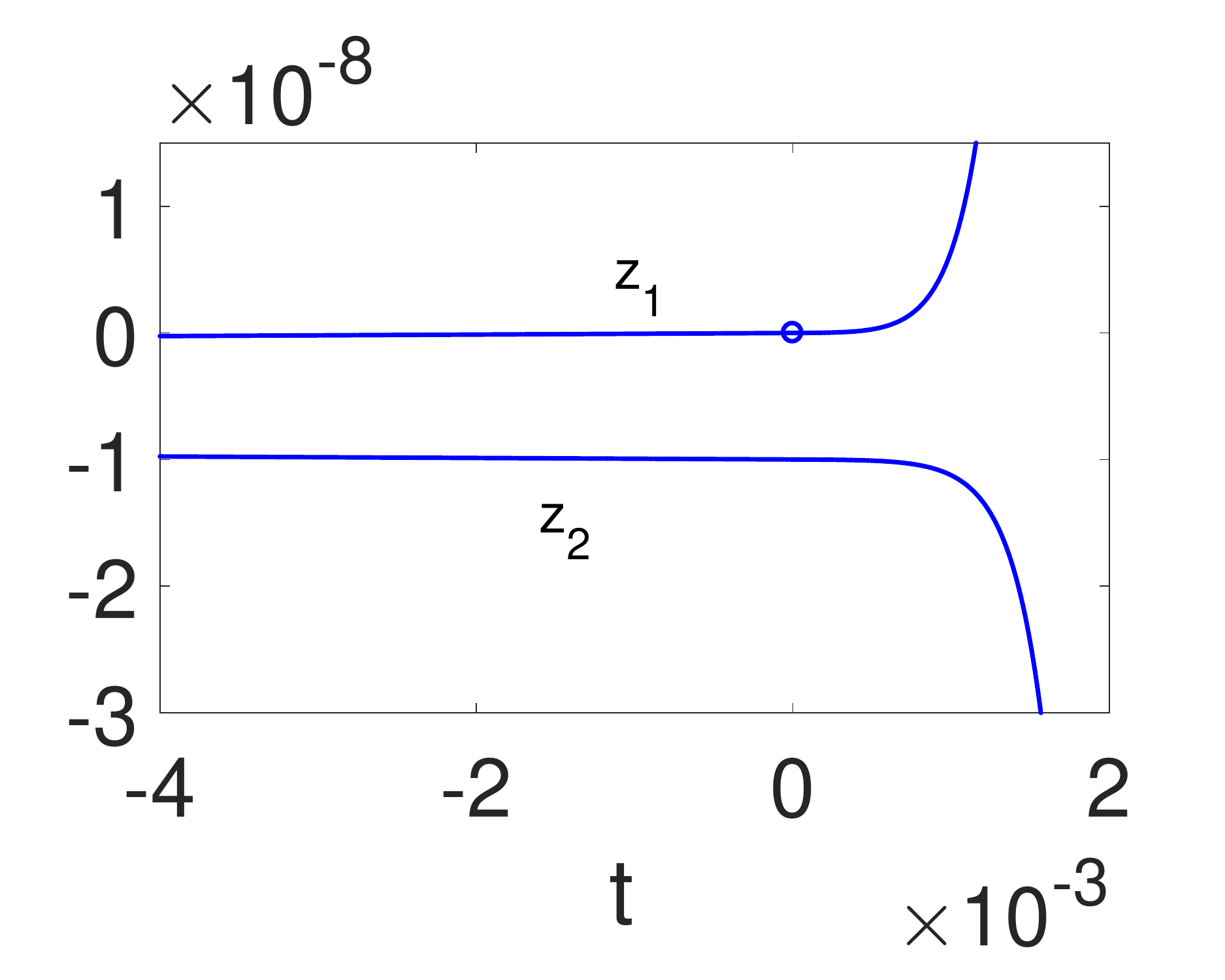}}
\caption{Crossing the L1 surface in numerical simulation of the model system introduced in Sec. \ref{sec:numericexample}.  L1 is crossed at $t=0$ in region 13 (a), 14 (b), or 46 (c). The diagrams show $z_1$(solid curve) and $z_2$ (dashed curve) versus time $t$. Circles mark the liftoff ($\lambda=0$) of the contact points. The corresponding model parameters and initial conditions are given in the appendix.
}
\label{fig:sim1}
\end{center}
\end{figure}

As we have seen, the contact force $\lambda_1$ becomes zero at L1, i.e. contact point 1 lifts off while point 2 continues to slip. We can now test the feasibility and stability of the FS mode. The result depends on where the L1 surface was crossed. Fig. \ref{fig:L1parts} shows a subdivision of L1 into regions marked by 2-digit numbers. The numbers refer to a classification scheme introduced by \citet{varkonyi2017dynamics}. The meaning of the labels is not explained here. Within ranges 23,12,13,41,42,31, the feasibility conditions \eqref{eq:FS1}-\eqref{eq:FS2} are satisfied, and we also have $p_{22}>0$, i.e. slip motion on contact 2 is feasible and stable (see Sec. \ref{sec:1contact-instantaneous}). Thus, we conclude that the system undergoes an SS$\rightarrow$FS contact mode transition. This scenario is illustrated by numerical simulation in Fig. \ref{fig:sim1}(a).

In contrast, if L1 is crossed in region 14 or 46, then we have $p_{22}<0<b_{22}$, implying that slip at contact point 2 is feasible but unstable. According to Table \ref{table:1contact}, the fast dynamics of point 2 may evolve either towards liftoff or towards an IWC (as long as contact 1 is separated). The unstable slip mode of point 2 corresponds to a saddle point of the fast contact dynamics, and its stable manifold separates the basins of attractions of liftoff and IWC. In order to determine, what happens to our system at this point, one should find the dynamic state of point 2 when point 1 lifts off to see within which basin of attraction it lies. Unfortunately, this simple approach is inconclusive here since it is straightforward to show that contact point 2 is exactly at the saddle point (which is part of the separatrix between liftoff and IWC). Hence, the analysis of the fast dynamics itself does not enable us to decide, in which direction the system evolves from here. The indeterminacy can perhaps be removed by considering the full slow-fast system. The full system evolves initially along the critical manifold corresponding to motion in SS mode. This manifold is probably off the separatrix between liftoff and IWC in the full slow-fast system. Nevertheless such a detailed analysis is beyond the scope of the present paper. Instead, we perform  numerical simulations using the system introduced in Sec. \ref{sec:numericexample}. Our results suggest evolution towards liftoff whenever the L1 surface is crossed in region 14, and towards IWC in region 46 (Fig. \ref{fig:sim1}(b)-(c)). Proving or refuting that the opposite scenarios (liftoff in region 46 and IWC in 14) are impossible remains an open question.

 In those two cases, which did occur in the simulations, the fate of the system can be deduced analytically:
\begin{itemize}
\item If point 2 evolves towards liftoff and we are in region 14, then the gradually decreasing contact force $\lambda_2$ and the negative signs of $p_{21},p_{22}$ imply that both contact points have increasing normal accelerations and eventually contact point 2 also lifts off. Then,  $b_{1},b_2>0$ imply that the FF mode persists, i.e. the system undergoes SS$\rightarrow$FF transition.
\item If point 2 evolves towards IWC, and we are in region 46, then $p_{21}>0$ implies that the large contact force  lifts up contact 1, i.e. it remains in separation. In this case an IWC develops at point 2, while point 1 is in F mode. IWC means that slip motion at point 2 stops rapidly, after which contact 2 may jump into separation with nonzero velocity
\end{itemize}  .


The results related to region 46 are particularly interesting, because Theorem \ref{thm: no_painleve_1_contact} and other results from \citet{champneys2016painleve} imply that a system with a single point contact never exhibits an IWC inside the \Pain regime ($P<0$). Nevertheless we see here that the analogous statement does not hold for systems with 2 point contacts. That is, we uncovered a generic mechanism, which does not occur in previously studied simple model systems.  

\begin{figure}
\centering
\subfigure[]{\includegraphics[width=0.49\textwidth]{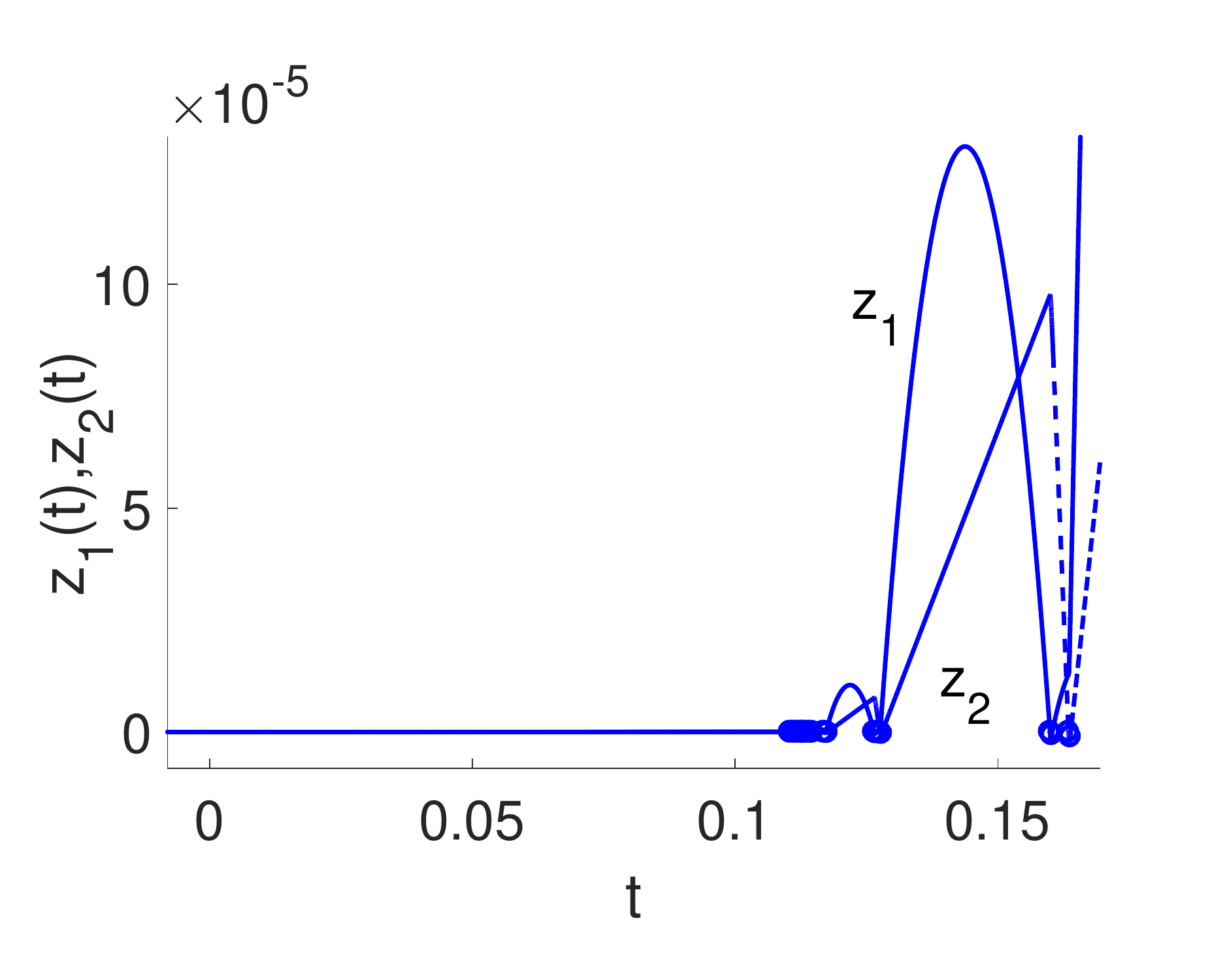}}
\subfigure[]{\includegraphics[width=0.49\textwidth]{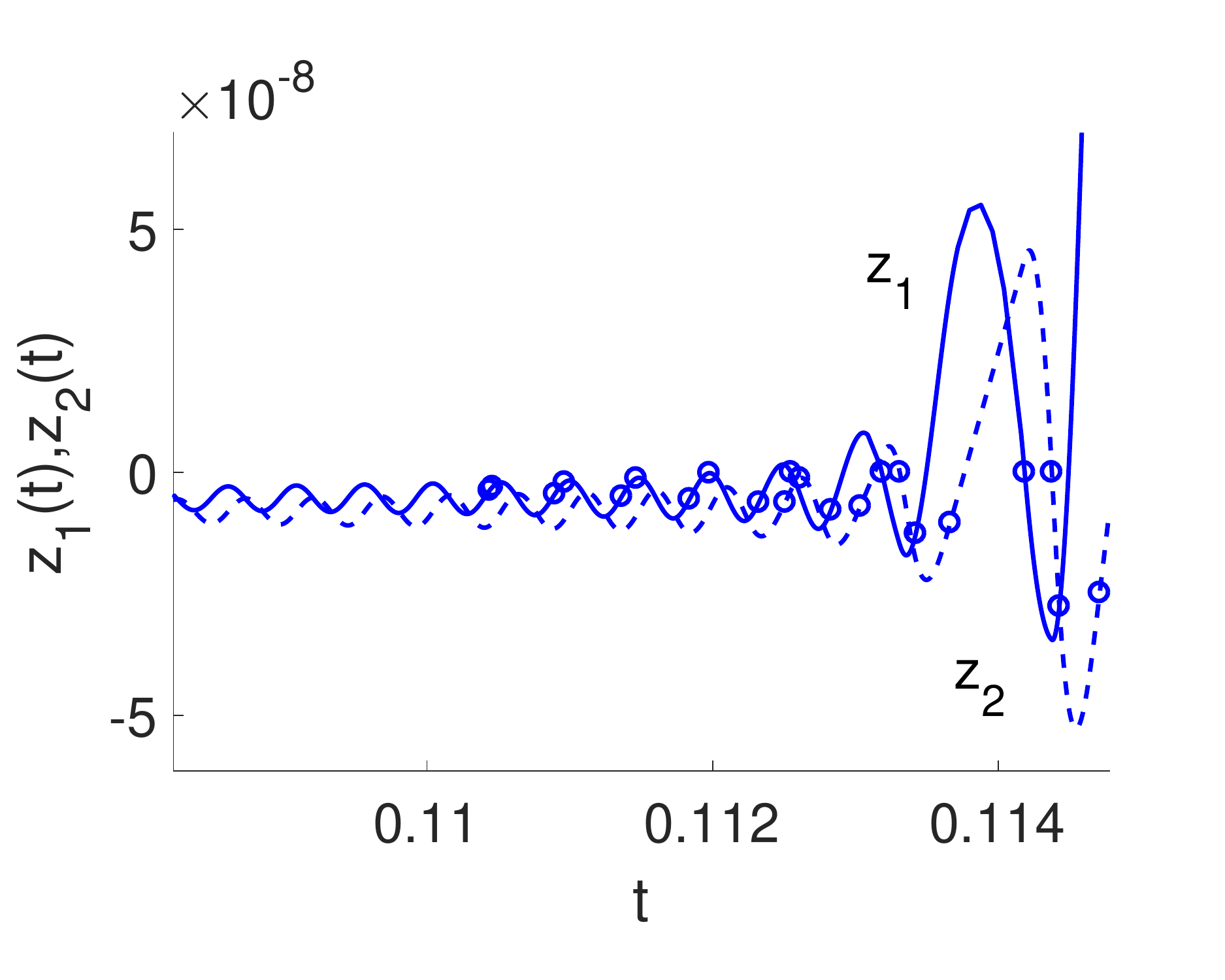}}
\subfigure[]{\includegraphics[width=0.49\textwidth]{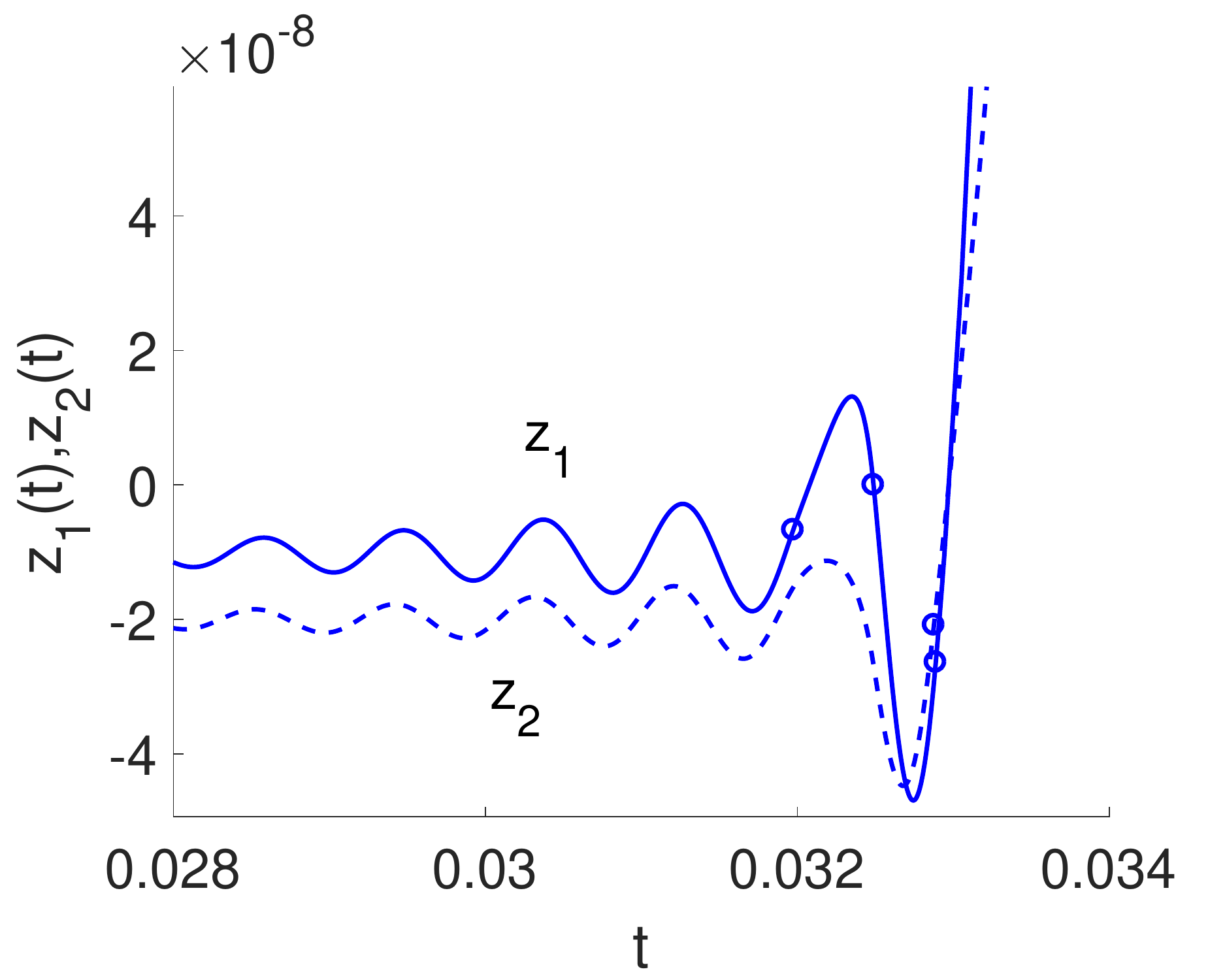}}
\subfigure[]{\includegraphics[width=0.49\textwidth]{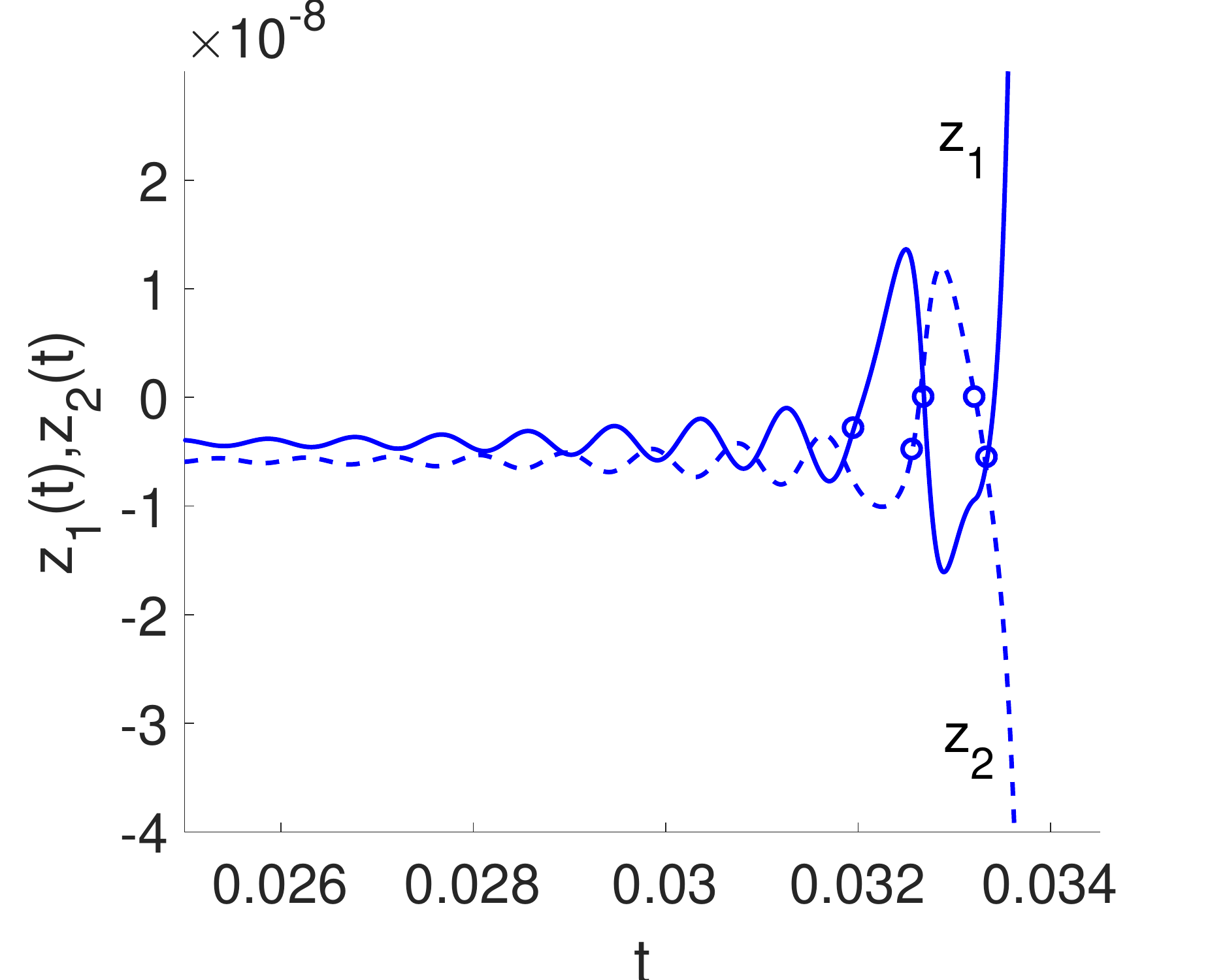}}
\subfigure[]{\includegraphics[width=0.49\textwidth]{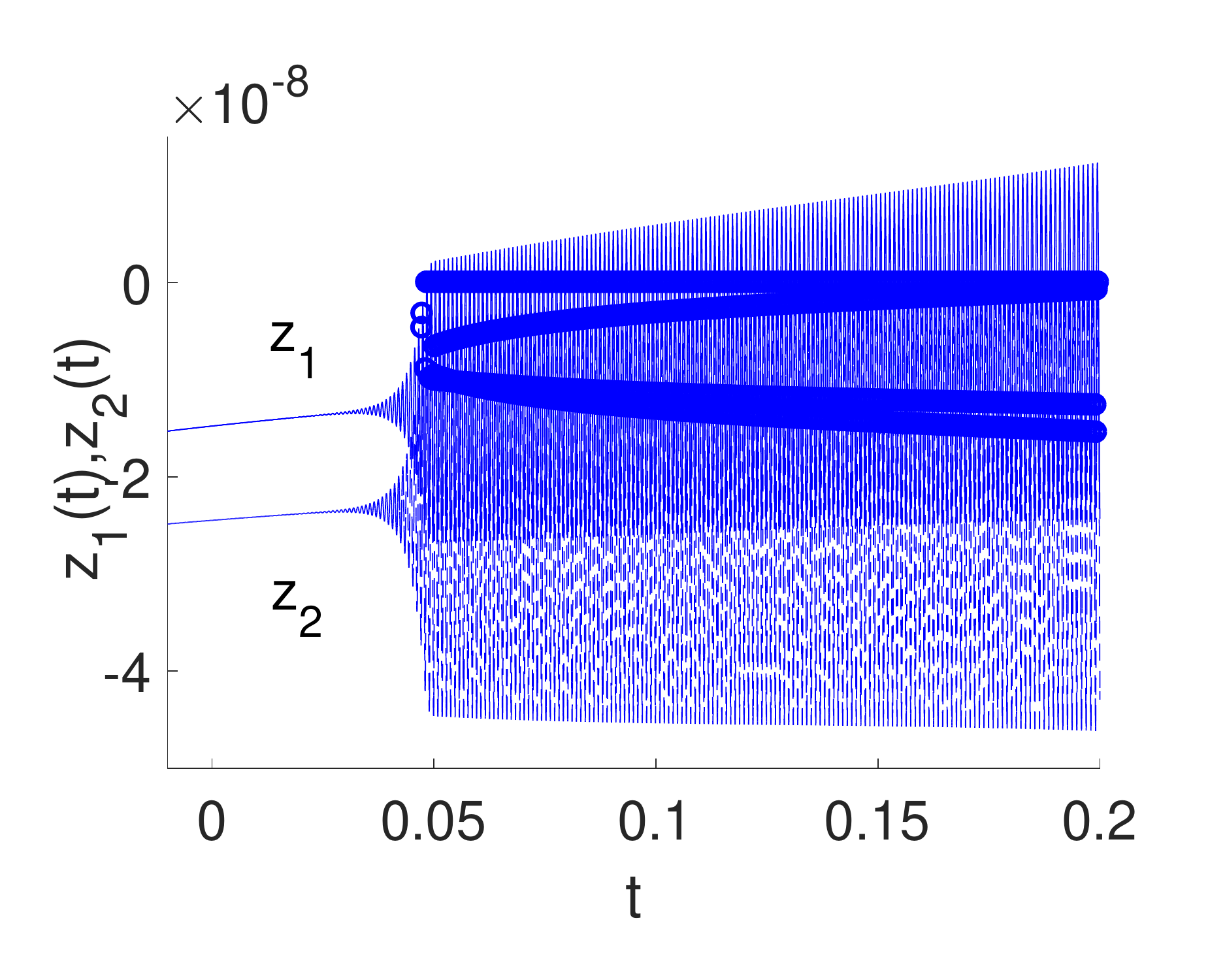}}
\subfigure[]{\includegraphics[width=0.49\textwidth]{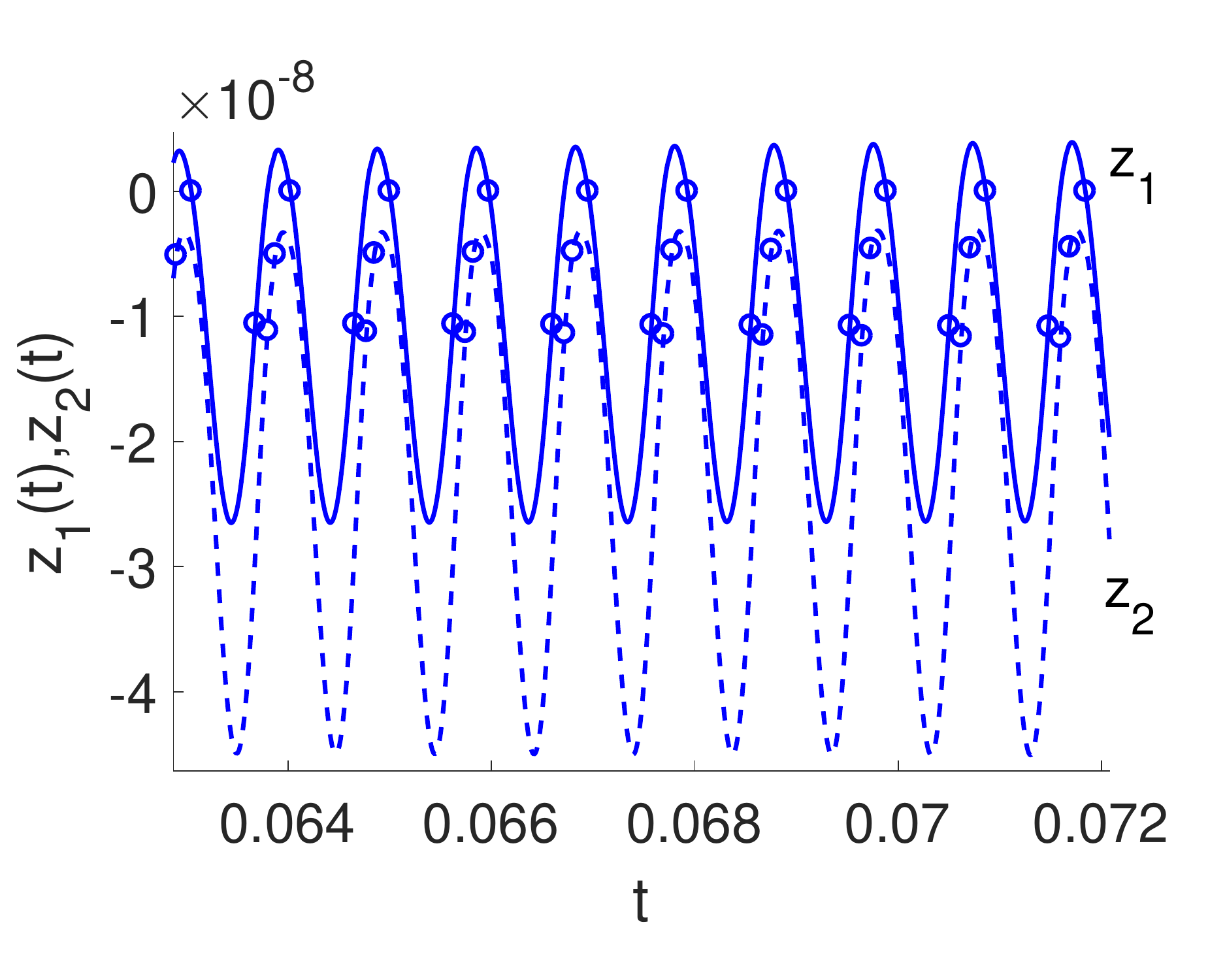}}
\caption{Crossing the S1 surface in numerical simulation of the model system.  Panel (a) shows inverse chattering  and (b) is a magnified detail of the same diagram. (c) is an example of simulatneous liftoff at both contact points, whereas (d) shows liftoff at point 1 accompanied by the onset of an IWC at point 2. (e) shows converges to a microscopic limit cycle of the fast subsystem and (f) is a magnified detail of the same diagram. 
}
\label{fig:sim2}
\end{figure}

\subsubsection{Crossing S1}
We have seen that the system matrix of \eqref{eq:contactdynamics_generaln} has a pair of purely imaginary eigenvalues at S1. As the system crosses the stability boundary, the invariant point of the regularized fast dynamics corresponding to the SS mode becomes unstable and $z_1$ and $z_2$ exhibit gradually growing harmonic oscillations. Soon, one of the two points will lift off. Beyond liftoff, the regularized contact law switches between the two cases of \eqref{eq:contactlaw} and it is not immediately clear what is going to happen. It can be shown that the FS and SF contact modes are either inconsistent or unstable, and FF is inconsistent except for a relatively small region within the S1 surface where $0<\beta<\pi/2$. During systematic numerical simulations, we found 4 different behaviors (Figure \ref{fig:sim2}): 
\begin{enumerate}
\item rapidly growing oscillations with repeated impacts and lift-off at both contact points ((Figure \ref{fig:sim2}(a-b))). The figure shows a regular oscillating pattern where the amplitude of the oscillation grows roughly by a factor of 10 in each cycle of the oscillation and thus only the last two cycles are visible. 
\item transition to FF mode ((Figure \ref{fig:sim2}(c))). The figure shows that $z_1$ and $z_2$ become positive and continue to grow rapidly.
\item IWC at point 1, while point 2 lifts off ((Figure \ref{fig:sim2}(d))) The figure shows that $z_1$ becomes positive and continues to grow rapidly, while $z_2$ simultaneously diverges towards minus infinity.
\item limit cycle of the fast dynamics involving repeated liftoff and reestablishment of the contacts. This behavior  tends to occur when $0<\gamma_2<\pi/2$ and $\beta+\pi$ is close to $\gamma_2$. The results depicted in Figure \ref{fig:sim2}(e-f) show oscillations of slowly growing amplitude (rather than exact limit cycles), since a finite value of $\epsilon$ was used in the simulation. 
\end{enumerate}
It is possible that other qualitatively different phenomena may also occur.
The first scenario listed above  appears to an external macroscopic observer as self-excited bouncing motion of increasing amplitude, i.e. inverse chattering \citep{paper2}. The last one appears as sliding combined with sustained microscopic, high-frequency vibration. Similar phenomena are known to lie behind brake squeal \citep{kinkaid2003automotive}.

\subsubsection{Crossing L2 or S2}
These scenarios are equivalent of the cases of crossing L1 and S1 with the only difference being that the roles of contact point 1 and 2 are reversed, furthermore the roles of regions 14 and 31, of 13 and 41, and of 23 and 46 are also reversed.

\subsubsection{Crossing P}
Matrix $P$ is singular along the surface P, hence \eqref{eq:lambda-2contact} shows that the contact forces diverge to infinity unless vector $b$ is a linear combination of the column vectors $p_i$ (which is satisfied at the GB1,GB2 lines). This property of surface P, and lines GB1, GB2 makes them similar to the \Pain manifold and the \GB manifold in the case of 1 point contact. This similarity suggests that a generalization of Theorem \ref{thm: no_painleve_1_contact} is true for 2 contacts as well. Indeed we will present a general theorem for system with arbitrary $n$ in Sec. \ref{sec:general n}, which implies that P is never reached away from the GB1, GB2 curves. The codimension 2 curves are investigated below.

\subsubsection{Attractive codimension 2 manifolds}

We have seen that the contact force has singularities at the codimension 2 manifolds L12, GB1, and GB2. In other words, the right-hand sides of the governing equations are discontinuous at these points. In such situations, the solution may be non-unique forward and/or backward in time. As a consequence, it is theoretically possible that any of these co-dimension 2 manifolds can be reached in finite time from an open set of initial conditions. In what follows, we construct conceptual examples to demonstrate that all of these manifolds may be reached from generic initial conditions, and we also investigate the consequences of such a transition.

\textbf{Crossing L12 (double liftoff singularity):} L12 is at the intersection of 2 liftoff boundaries, which inspires the name proposed for this phenomenon. We consider the system \eqref{eq:dotP}-\eqref{eq:dotb} with ideally rigid contacts. Then, by \eqref{eq:lambda-2contact} the system becomes
\begin{align}
\dot P = A_1\label{eq:dotP2}\\
\dot b =  \alpha_2- A_3P^{-1}b \label{eq:dotb2}
\end{align}
The dynamics of $P$ is determined uniquely by the choice of
\begin{align}
A_1=\left[
\begin{matrix}
-1&1\\
0&0
\end{matrix}
\right]
,P_0=\left[
\begin{matrix}
0&0\\
1&1
\end{matrix}
\right]
\label{eq:alpha1P0}
\end{align}
This choice implies that the angle $\gamma_1$ grows while $\gamma_2$ decreases and both angles become equal to $\pi/2$ at $t=0$. 
In addition, we choose the values
\begin{align}
\alpha_2=
\begin{bmatrix}
0\\
0
\end{bmatrix}
,A_3=\left[
\begin{matrix}
2&-2\\
0&0
\end{matrix}
\right]
\label{eq:alpha23}
\end{align}

Assume now that the system has an initial state at some $t_{init}<0$ where  the system is initially in SS mode and the system parameters are within the region of feasibility and stability of SS, i.e.:
\begin{align}
\gamma_1<\beta+\pi<\gamma_2
\label{eq:SSfeasible-stable}
\end{align}
where $\beta$ and $\gamma_i$ are the angles illustrated by Fig. \ref{fig:gammabeta}. As we approach $t=0$, the gap between the $\gamma_1$ and $\gamma_2$ angles shrinks to 0. Nevertheless,
\begin{lemma}
 For any initial condition satisfying \eqref{eq:SSfeasible-stable} at $t_{init}<0$, the same condition is not violated as long as $t<0$ and thus the system remains in SS mode until reaching the L12 manifold at $t=0$. 
 \label{lem:L12}
\end{lemma}
\begin{figure}[h]
\begin{center}
\includegraphics [width=10cm] {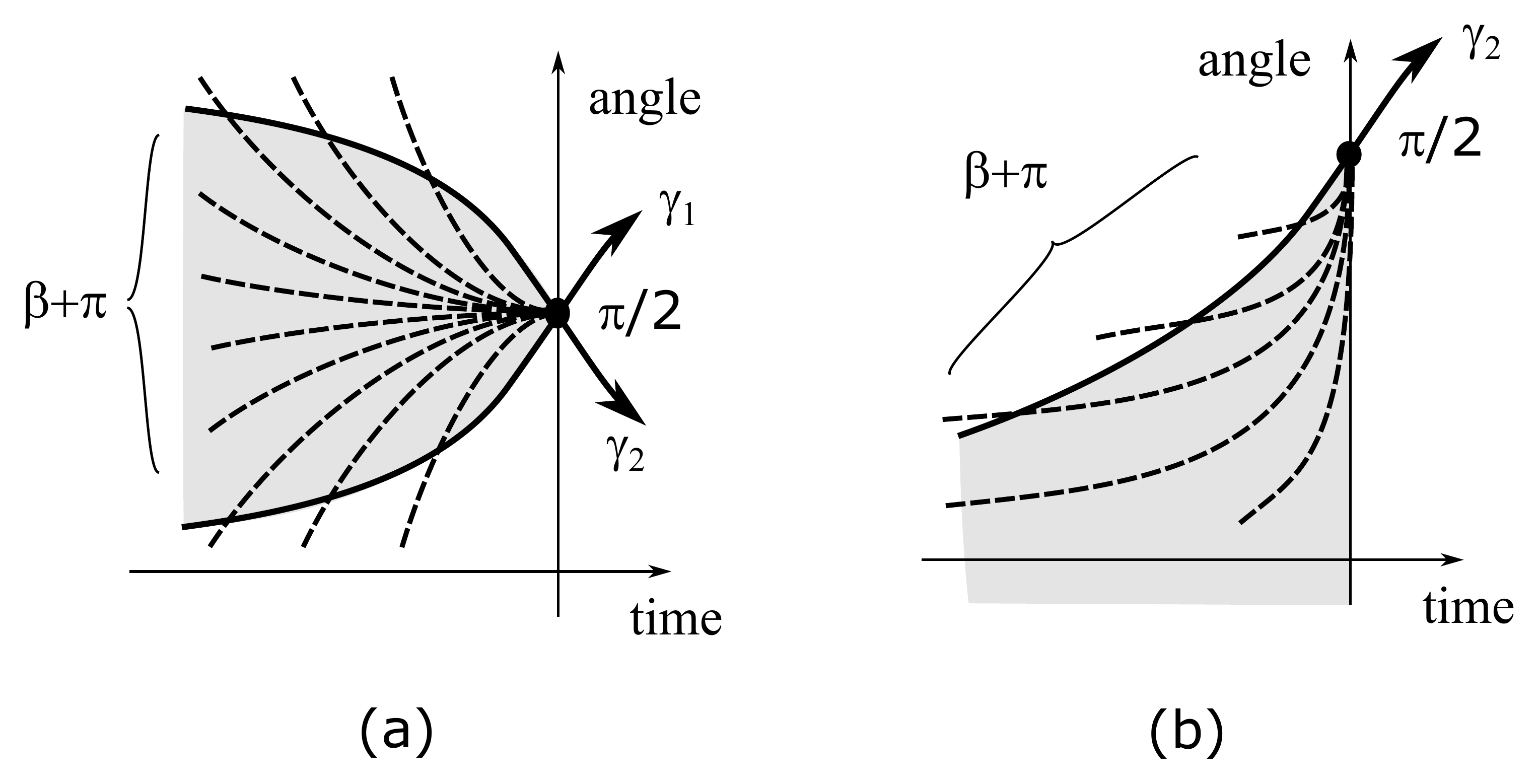}
\caption{Illustrations of the statements of Lemma \ref{lem:L12} (a) and Lemma \ref{lem:GB1} (b). Solid curves indicate the evolution of the angles $\gamma_i$. The dashed curves show the evolution of the angle $\beta+\pi$ for various initial conditions. The SS mode is feasible and stable in the grey areas. Filled circles mark the attractive singular points.}
\label{fig:lemmas}
\end{center}
\end{figure}
In the proof we show that whenever $\beta=\gamma_1$, the resulting contact force causes $\beta$ to increase faster then $\gamma_1$, whereas in the case of $\beta=\gamma_2$, the resulting contact force makes $\beta$ decrease faster than $\gamma_2$. This mechanism keeps $\beta$ in the shrinking interval $(\gamma_1,\gamma_2)$ until the L12 manifold is reached at $t=0$ (Fig. \ref{fig:lemmas}(a)). The detailed proof, given in the appendix, borrows ideas from  \citet{Genot1999}. We apply a singular rescaling of time and show that the L12 manifold turns into an attractive invariant point of the rescaled dynamics. We note that the proof of Lemma \ref{lem:L12} does not rely on the specific values of $A_1,A_3,\alpha_2$ chosen above, hence the mechanism illustrated by this example may occur in many systems.

\begin{figure}[h]
\begin{center}
\includegraphics [width=6.5cm] {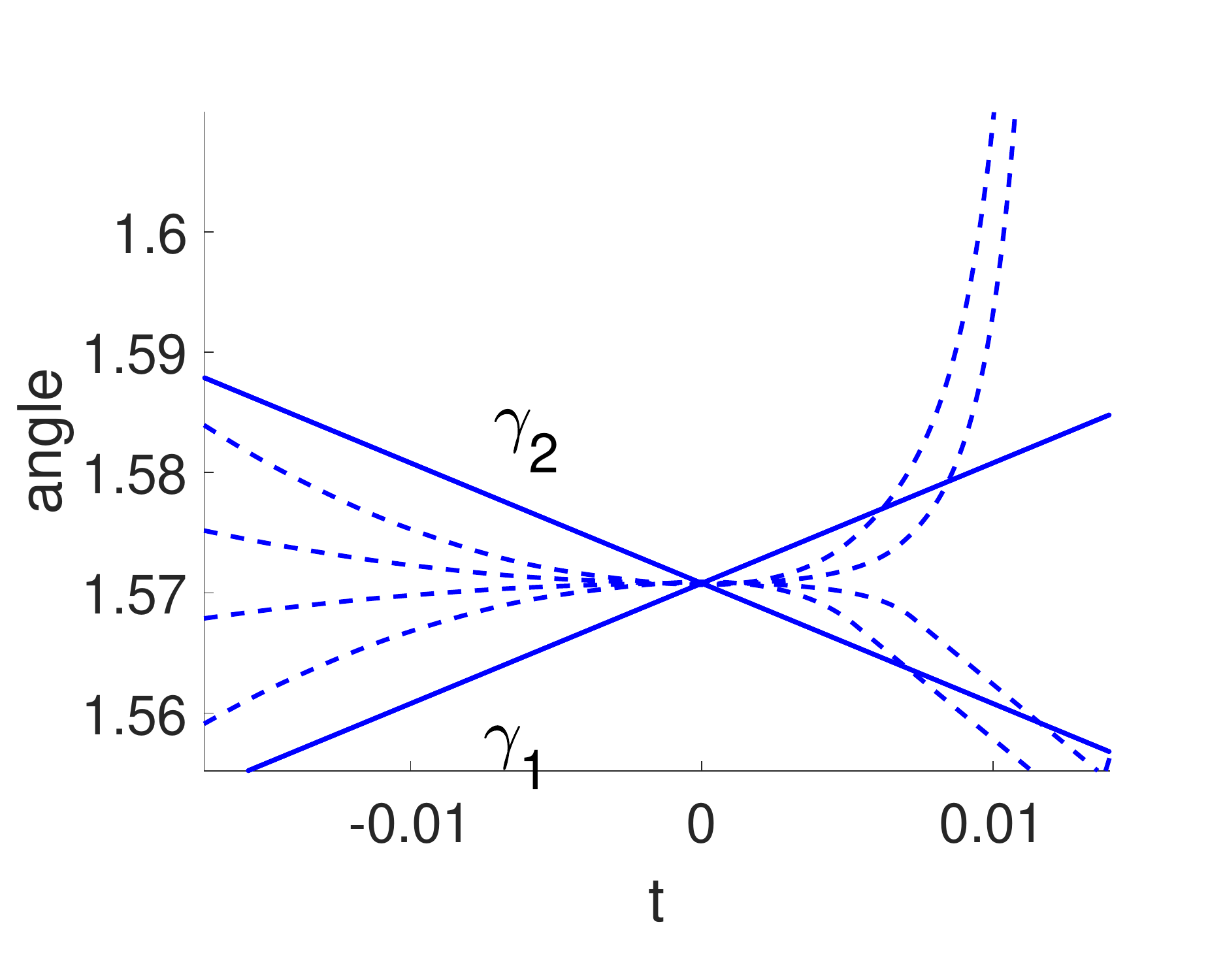}
\includegraphics [width=6.5cm] {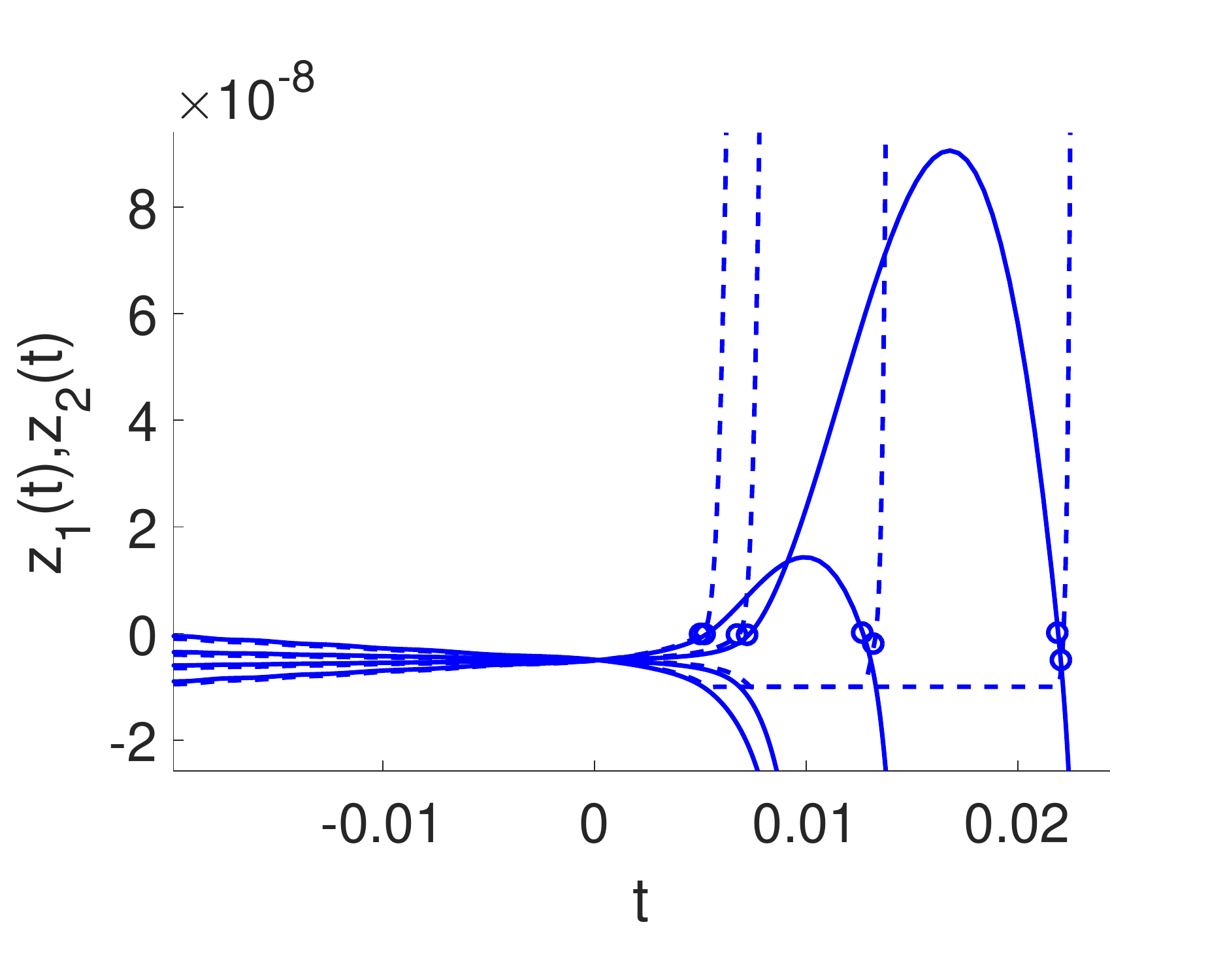}
\caption{Numerical simulation of the double liftoff singularity. Left: $\beta+\pi$ (dashed curves) and $\gamma_i$ (solid curves) versus time for four different initial conditions. Right: $z_1$ (solid curve) and $z_2$ (dashed curves) in the same simulations. 	
}
\label{fig:sim3}
\end{center}
\end{figure}

At the L12 manifold, the contact force $\lambda$ becomes undefined due to the singularity of $P$, and the dynamics of $b$ given by \eqref{eq:dotb2} becomes ill-defined. The indeterminacy may be resolvable by contact regularization, nevertheless such an analysis is beyond the scope of the present work. For the sake of illustration, we show examples of numerical simulation using regularized contact in Fig. \ref{fig:sim3}. The results confirm that angle $\beta$ remains in the shrinking interval $(\gamma_1,\gamma_2)$ until the singularity is reached. After crossing the singularity contact 1 lifts off while 2 remains in slip state for some initial conditions. Shortly thereafter, contact 1 becomes active again, and it exhibits rapidly increasing contact force (indicating the onset of an IWC), while contact 2 lifts off. For other initial conditions, contact 2 lifts off and contact one enters an IWC directly. This behavior appears to be qualitatively similar to one of the possible scenarios during the dynamic jamming singularity of the slipping rod reported by \citet{nordmark2017dynamics}.

\textbf{Crossing GB1 (dynamic jam):} we again consider the system introduced in Sec. \ref{sec:numericexample} with rigid contacts. This time, we choose 
\begin{align}
A_1=\left[
\begin{matrix}
-1&-1\\
0&0
\end{matrix}
\right]
,P_0=\left[
\begin{matrix}
0&0\\
-1&1
\end{matrix}
\right]
,
\alpha_2=
\begin{bmatrix}
0\\
0
\end{bmatrix}
,
A_3=\left[
\begin{matrix}
0.5&0.5\\
0&0
\end{matrix}
\right]
\label{eq:alpha1P0masodik}
\end{align}
Then, the angle $\gamma_1$ decreases while $\gamma_2$ increases, and we have $\gamma_1=\gamma_2+\pi=3\pi/2$ at $t=0$. Furthermore,
\begin{lemma}
 For any initial condition such that the system is in SS mode at time $t_{init}<0$ and $0<\beta+\pi<\gamma_2$, the SS mode persists until $t=0$, where the angle $\beta$ converges to $3\pi/2$, i.e. the system reaches the GB1 manifold.
 \label{lem:GB1}
 \end{lemma}
The proof is based on the observation that large $\lambda$ causes $b_1$ to converge towards 0 rapidly (Fig. \ref{fig:lemmas}(b)). Technically, the proof is very similar to that of Lemma \ref{lem:L12} and it is again given in the appendix.

\begin{figure}[h]
\begin{center}
\includegraphics [width=6.5cm] {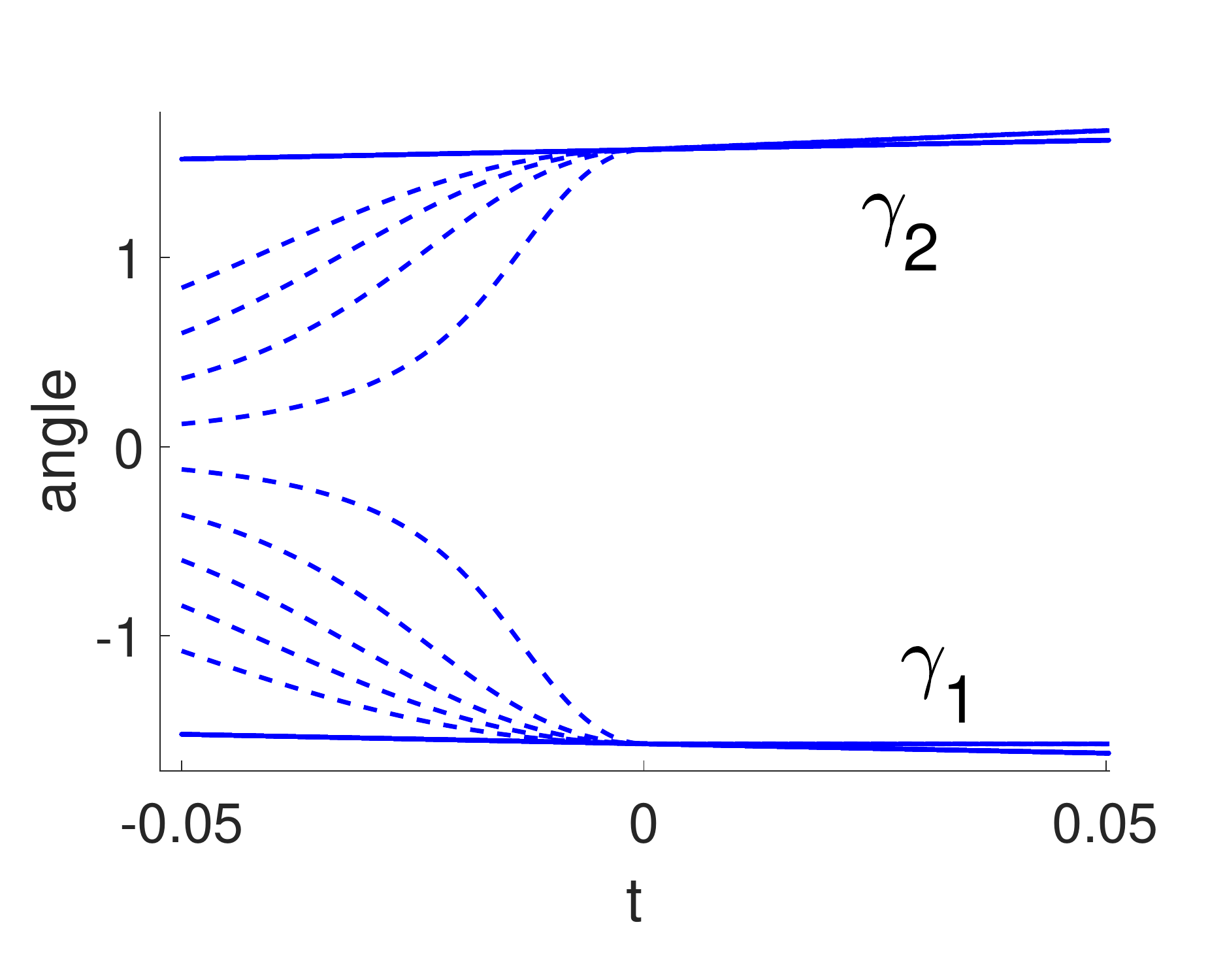}
\includegraphics [width=6.5cm] {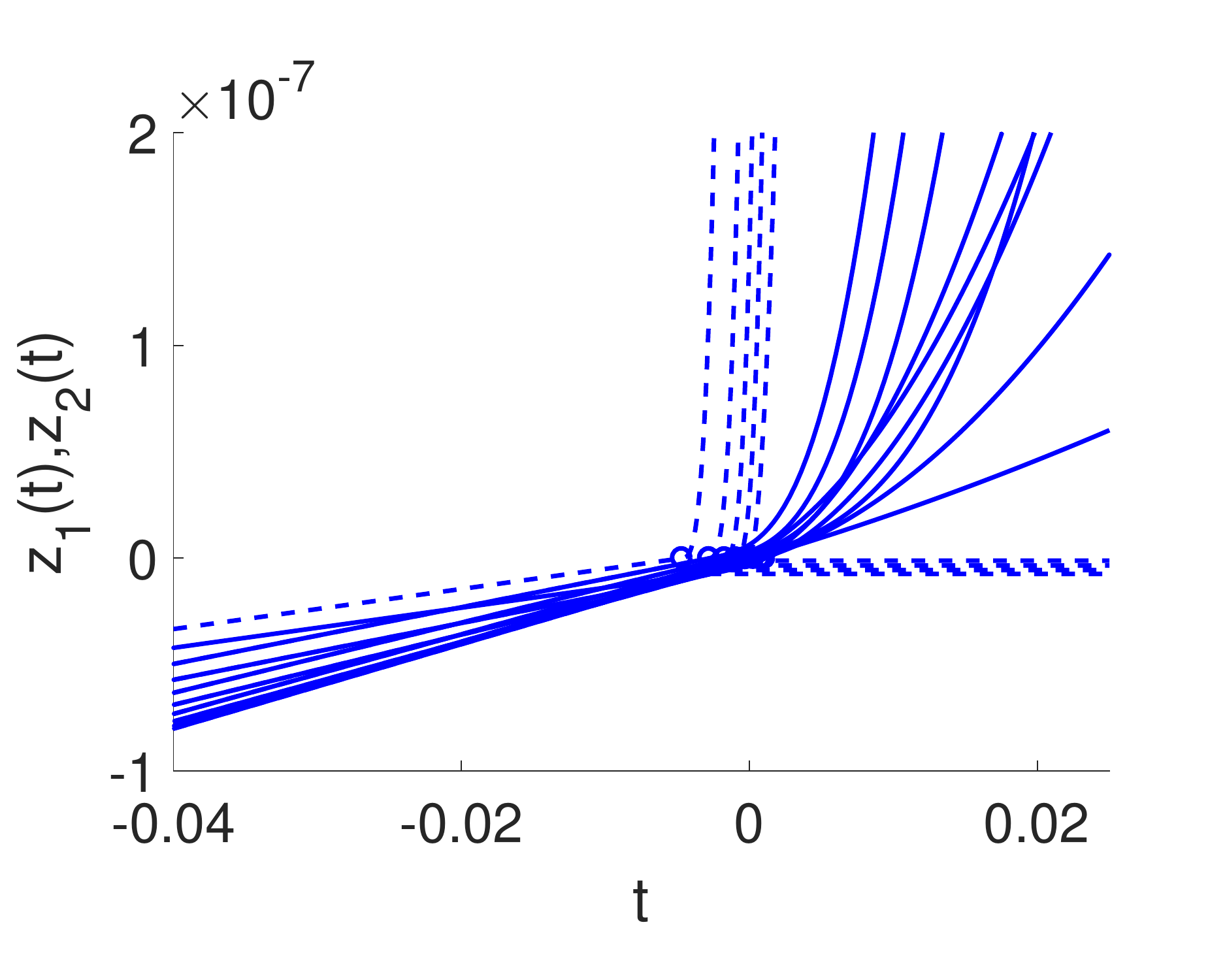}
\caption{Numerical simulation of dynamic jamming. Left: $\beta+\pi$ (dashed curves) and $\gamma_i$ (solid curves) versus time for 8 different initial conditions. Right: $z_1$ (solid curves) and $z_2$ (dashed curves) in the same simulations.
}
\label{fig:sim4}
\end{center}
\end{figure}

Upon reaching the GB1 or GB2 manifold, the contact force again becomes ill-defined due to the singularity of $P$. Systematic investigations of what happens after this point is beyond the scope of the paper. Based on lessons learned from the single-contact case \citep{nordmark2017dynamics,kristiansen2017canard}, we expect that the regularized system exhibits liftoff, or impulsive contact forces. Numerical simulations (Fig. \ref{fig:sim4}) are consistent with our findings and expectations. Depending on the initial conditions, solution trajectories converge to $\beta=\pi/2$ or $3\pi/2$ at $t=0$, i.e. the system is attracted by the GB1 or by the GB2 manifold.In the first case, contact 1 lifts off, whereas both ones lift off in the second. Further simulations (not shown) indicated that impulsive contact forces are possible, furthermore the initial liftoff is often followed by additional mode transitions, similarly to the results of Fig. \ref{fig:sim3} and to numerical results of \citet{nordmark2017dynamics} in the case of $n=1$.  

\subsection{Dynamic jamming of general systems} \label{sec:general n}

We have seen previously that the boundary of the region of feasibility and stability of slip motion includes the codimension-1  \Pain surfaces where $P=0$ ($n=1$) or $detP=0$ ($n=2$). Nevertheless general points of these surfaces are inpenetrable to the systems. In addition, we have also seen that some codimension-2 surfaces with $detP=0$ may become attractive in turn. In what follows, we generalize Theorem \ref{thm: no_painleve_1_contact} to systems with arbitrary number of point contacts in two dimensions and thus show that the inpenetrability of the \Pain surfaces is a general property in contact mechanics of planar systems. 

Assume that a mechanical system  with $n$  point contacts initially undergoes slip motion at all contact points with $\dot x_i>0$. Similarly to the case $n=2$, slip motion must be feasible, which means that $-b$ must be in the cone spanned by the column vectors of $P$.

First we define the \Pain manifold as follows:
\begin{definition}
 The \Pain manifold in the state space consists of those points where $P$ is singular and the cone spanned by column vectors of $P$ is a full $(n-1)$-dimensional space.
\end{definition}

If we approach a generic point of the \Pain manifold in state space transversally, the cone spanned by the column vectors of $P$ gradually grows and converges to an $n$-dimensional half-space. When crossing the manifold, the cone collapses to an $n-1$ dimensional space and then it turns inside out, i.e. it becomes a half-space again. Because of this property, slip motion is feasible on exactly one side of the \Pain manifold, hence the \Pain manifold may form a boundary of the range of feasible and stable slip motion in state space.

Next, we define the \GB manifold as follows:
\begin{definition}
The \GB manifold is a submanifold of the Painlev\'e manifold including those points where $b$ is contained in the $n-1$ dimensional space spanned by the column vectors of $P$.
 \end{definition}
It is easy to show that by approaching the \GB manifold in a direction transversal to the \Pain manifold, one  of the contact forces in slip mode dictated by \eqref{eq:lambda-1contact}  converges to 0. Hence, the \GB manifold contains those points within the \Pain manifold, which are at the verge of liftoff (in full analogy with the $n=1$ case).

Now we can state the following theorem: 
\begin{theorem}
If $P$ is Lipschitz with respect to $q$ and $a,b,K$ are continuous function of their arguments, then the system never reaches the Painlev\'e manifold except for points of the \GB manifold. 
\label{thm: no_painleve_n_contact}
\end{theorem}

The proof is highly similar to that of Theorem \ref{thm: no_painleve_1_contact} but involves some additional technical steps. We present the proof in the Appendix.

\section{Conclusions}

In this paper, we investigated singularities and transitions of mechanical systems with frictional point contacts during slip motion. We found that the presence of 2 (or more) point contacts induces several dynamic phenomena, which are not possible in previously studied single-contact systems. Slip motion may terminate due to destabilization or loss of feasibility. In both cases, various transitions may occur, including lift-off at one or both contact points, self-excited microscopic limit cycle oscillations or exponentially growing macroscopic oscillations, as well as impact without collisions. The last three transitions are not possible in the single-contact case. The list of phenomena is not comprehensive, as other types of behavior may be possible. On the other hand, the author believes that the patterns found by this study occur frequently in systems with any number of contacts and thus they are of practical interest.

We also demonstrated that two different types of singularity are generic in the case of $n=2$. The first one occurs when a system converges to a codimension-2 manifold, where the associated matrix $P$ is singular and simultaneously the system is at the boundary of lift-off at one contact point.  This singularity is essentially identical to the previously uncovered \emph{dynamic jamming} singularity of single-contact systems first described by \citet{Genot1999}. Based on earlier results, we expect that the contact force may or may not diverge to infinity, and passing the singularity may be followed by impulsive contact forces or lift-off at one point. A second type of singularity occurs when the system converges to another codimension-2 manifold, which is at the boundary of liftoff at both contact points. The double liftoff singularity is a novel phenomenon. Here, the contact force does not diverge to infinity, nevertheless the system shows indeterminacy at the point of passing the singularity. Either one of the contact points may lift off, while the other contact may continue to slip or exhibit impulsive contact forces. The underlying mechanism and the induced indeterminacy are both similar to the well-known two-fold singularity of Filippov system \citep{Springerbook}.

The analysis was based on the assumption of a picewise linear, regularized contact model with two parameters ($k_i,\nu_i$). The phenomena uncovered in the paper did not rely on choosing special values of the parameters $k_i,\nu_i$ of the model, hence they appear to be generic. The assumption of linearity was motivated by its simplicity, but it is not crucial either. In the case of a nonlinear contact model (such as Hertz law), the stability of contact modes can be investigated after linearization, and the result of the analysis is often independent of the presence or the exact form of nonlinearity \citep{champneys2016painleve}. Hence we believe that the uncovered phenomena also occur in systems with nonlinear contacts.  

This paper leaves many questions open for future work. Our aim was to provide a general overview of generic transitions, and we skipped the detailed analysis of these transitions. Open questions include
\begin{itemize}
\item detailed description of the dynamics after crossing the liftoff manifold within region 14 and 46
\item comprehensive list of transitions for $n=2$ (or more) contacts after crossing the stability boundaries and conditions under which they occur
\item general characterization of the double lift-off and dynamic jamming singularities for nonlinear systems with $n=2$ (or more) contacts
\end{itemize}
Also, we did not present real examples of the most interesting transitions and singularities. It was demonstrated that these singularities occur in a conceptual model, nevertheless finding real-world examples will be subject of future work.

\begin{acknowledgement}
The author thanks two anonymous reviewers for their useful suggestions. This work was supported by Grant 124002 of the National Research, Development, and Innovation Office of Hungary.
\end{acknowledgement}

\appendix
\section{Parameters and initial conditions of the numerical simulation}

In all simulations, we use $\epsilon=10^{-4}$, $k_1=k_2=\nu_1=\nu_2=1$. In the first set of simulations (Fig. \ref{fig:sim1}, \ref{fig:sim2}), we have
$$
\alpha_2=
\begin{bmatrix}
0\\0
\end{bmatrix}
,
A_3=
\begin{bmatrix}
0&0\\0&0
\end{bmatrix}
$$
which means that $b$ is constant throughout the simulation. The values of parameter $A_1$  are shown in Table \ref{table:parameters}. The simulations are launched at $t_{init}=-0.01$ (which allows initial transients to relax before $t=0$), and the initial conditions of $P$ and $b$ are chosen in such a way that transitions from slip occur exactly at $t=0$. We show in Table \ref{table:parameters} the critical values $P_0$ and $b_0$ at $t=0$, which determine initial values $P_{init}$ and $b_{init}$ uniquely. The remaining initial conditions are $\dot \dot z_{init}=0$ and $z_{init}=-P_{init}^{-1}b$, which corresponds to the stationary values during sustained slip motion. 

In a second set of simulations, we illustrated that systems may reach the codimension 2 singular manifolds (Fig. \ref{fig:sim3}-\ref{fig:sim4}). Here we chose the same values of $\epsilon$, $k_i$,$\nu_i$, and $\alpha_2$ as before. We used the values of $P_0$, $A_1,A_3$ given by \eqref{eq:alpha1P0}, \eqref{eq:alpha23}, \eqref{eq:alpha1P0masodik}, $t_{init}=-0.05$ and several initial conditions $b_{init}$.
 
\begin{table}
\centering
\caption{Parameter values and initial conditions of the numerical simulations}
\label{table:parameters}
\begin{tabular}{cp{25mm}ccc}
\\
\hline
             Figure number & event at $t=0$ & $P_0$ & $A_1$ & $b_0$  \\
\hline
Fig. \ref{fig:sim1}(a)    & crossing L1\newline transition to FS &
$\begin{bmatrix}
\cos(0.48\pi)&\cos(0.52\pi)\\ \sin(0.48\pi)&\sin(0.52\pi)\end{bmatrix}$    
&  
$\begin{bmatrix}
0&1\\ 0&0\end{bmatrix}$         
&  
$\begin{bmatrix}
\cos(1.52\pi)\\ \sin(1.52\pi)\end{bmatrix}$  \\\\
Fig. \ref{fig:sim1}(b)    & crossing L1\newline transition to FF &
$\begin{bmatrix}
\cos(0.1\pi)&\cos(1.05\pi)\\ \sin(0.1\pi)&\sin(1.05\pi)\end{bmatrix}$    
&  
$\begin{bmatrix}
0&0\\ 0&1\end{bmatrix}$         
&  
$\begin{bmatrix}
\cos(0.05\pi)\\ \sin(0.05\pi)\end{bmatrix}$  \\\\
Fig. \ref{fig:sim1}(c)    & crossing L1\newline transition to IWC &
$\begin{bmatrix}
\cos(1.9\pi)&\cos(1.95\pi)\\ \sin(1.9\pi)&\sin(1.95\pi)\end{bmatrix}$    
&  
$\begin{bmatrix}
0&0\\ 0&-1\end{bmatrix}$         
&  
$\begin{bmatrix}
\cos(0.95\pi)\\ \sin(0.95\pi)\end{bmatrix}$  \\\\
Fig. \ref{fig:sim2}(a,b)    & crossing S1\newline inverse chattering&
$\begin{bmatrix}
\cos(1.7240\pi)&\cos(0.2\pi)\\ \sin(1.7240\pi)&\sin(0.2\pi)\end{bmatrix}$    
&  
$\begin{bmatrix}
0&0\\ -1&0\end{bmatrix}$         
&  
$\begin{bmatrix}
-1\\ 0\end{bmatrix}$  \\\\
Fig. \ref{fig:sim2}(c)    &  crossing S1\newline transition to FF &
$\begin{bmatrix}
\cos(0.1403\pi)&\cos(1.1\pi)\\ \sin(0.1403\pi)&\sin(1.1\pi)\end{bmatrix}$    
&  
$\begin{bmatrix}
0&-1\\ 0&0\end{bmatrix}$         
&  
$\begin{bmatrix}
\cos(0.05\pi)\\ \sin(0.05\pi)\end{bmatrix}$  \\\\
Fig. \ref{fig:sim2}(d)    & crossing S1\newline IWC &
$\begin{bmatrix}
\cos(1.8597\pi)&\cos(1.9\pi)\\ \sin(1.8597\pi)&\sin(1.9\pi)\end{bmatrix}$    
&  
$\begin{bmatrix}
0&0\\ -1&0\end{bmatrix}$         
&  
$\begin{bmatrix}
\cos(0.88\pi)\\ \sin(0.88\pi)\end{bmatrix}$  \\\\
Fig. \ref{fig:sim2}(e,f)    & crossing S1\newline microscopic oscillation&
$\begin{bmatrix}
\cos(1.4\pi)&\cos(0.3597\pi)\\ \sin(1.4\pi)&\sin(0.3597\pi)\end{bmatrix}$    
&  
$\begin{bmatrix}
0&0\\ -1&0\end{bmatrix}$         
&  
$\begin{bmatrix}
\cos(1.3\pi)\\ \sin(1.3\pi)\end{bmatrix}$  
\\
\hline
\end{tabular}
\end{table}

\section{Proof of Lemma \ref{lem:L12}}
Let us introduce the notation $\overline\gamma_i=\gamma_i-\pi/2$, $\overline\beta=\beta+\pi/2$. In the new variables, the feasibility and stability of the SS mode requires $\overline\gamma_2>\overline\beta>\overline{\gamma}_1$, the L1 manifold is given by $\overline\gamma_2=\overline\beta$ and the L12 manifold by $\overline\gamma_1=\overline\gamma_2=\overline\beta=0$.  

From \eqref{eq:alpha1P0} we deduce 
\begin{align}
P=
\begin{bmatrix}
-t&t\\1&1
\end{bmatrix}
\end{align}
implying
\begin{align}
\tan(\overline\gamma_2)&=-t\\
\frac{d}{dt}(\tan\overline\gamma_2)&=-1
\label{eq:dotgamma1}\\
(\tan\overline\gamma_2)'&=t=-\tan\overline\gamma_2
\label{eq:gamma1'}
\end{align}
where $'$ denotes differentiation with respect to singularly rescaled time $s$, given by $d/dt=|t|\cdot d/ds$.

At the same time the equations \eqref{eq:alpha23}, \eqref{eq:dotb2} yield
\begin{align}
\dot b=
\begin{bmatrix}
2t^{-1}&0\\0&0
\end{bmatrix}b
\end{align}
Hence, $b_2$ is constant and the derivative of $b_1$ with respect to rescaled time can be expressed as
\begin{align}
b_1'=
-2b_1
\end{align}
This equation can be rearranged as
\begin{align}
(b_1/b_2)'=
-2(b_1/b_2)
\\
(\tan\overline\beta)'=
-2\tan(\overline\beta)
\label{eq:beta'}
\end{align}

Equations \eqref{eq:gamma1'} and \eqref{eq:beta'} show that the point $\overline{\beta}=\overline{\gamma_2}=0$ is an attractive invariant point of the singularly rescaled dynamics, furthermore whenever $\overline{\beta}=\overline{\gamma_2}$, then the variable $\overline{\beta}$ approaches 0 faster than $\overline{\gamma_2}$. Hence trajectories of the dynamics never cross the L1 surface inside out. It can be shown in a similar fashion that trajectories never cross the L2 surface inside out, which completes the proof. 

\section{Proof of Lemma \ref{lem:GB1}}
We again introduce $\overline\gamma_2$ and $\overline\beta$ as in the proof of Lemma \ref{lem:L12}. Then, the SS mode is feasible and stable as long as $0>\overline\gamma_2>\overline\beta$,  the L1 surface is given by $\overline\gamma_2=\overline\beta$, and the GB1 manifold by $\overline\gamma_2=\overline\beta=0$.  By applying the same rescaling of time, we arrive to a pair of equations similar to  \eqref{eq:gamma1'} and \eqref{eq:beta'}:
\begin{align}
(\tan\overline\gamma_2)'&=-\tan\overline\gamma_2\\
(\tan\overline\beta)'&=
-0.5\tan\overline\beta+O(t^2)
\end{align}
We can conclude again, that $\overline{\beta}=\overline{\gamma_2}=0$ is an attractive fixed point. Nevertheless in this case, $\overline{\beta}$ converges to 0 slower than $\overline{\gamma_2}$ whenever they are equal, which means that the L1 surface us never crossed by trajectories from inside out, and the SS mode persists until the GB1 manifold is reached.

\section{Proof of Theorem \ref{thm: no_painleve_n_contact}}

Assume that the system is launched at $t=t_{init}<0$ and a state on the \Pain manifold but away from the \GB manifold at time $t=0$. 

The contact force given by \eqref{eq:lambda-2contact} can be expressed as 
\begin{align}
\lambda&=-det(P)^{-1}
adj(P)b
\\
&:=det(P)^{-1}\overline\lambda
\label{eq:lambda P}
\end{align}
where $adj(P)$ is the adjugate matrix of $P$. 
Being on the \Pain manifold implies that
\begin{align}
det(P_0)=0
\label{eq:detP=0}
\end{align}
Being on the \GB manifold would mean that despite the singularity of $P_0$, there exists a scalar $\lambda$ satisfying $P_0\lambda+b_0=0$. Then this equation can be multiplied by $adjP_0$ on the left to obtain 
$$
\underbrace{adjP_0P_0}_{=detP_0I=0}\lambda+\underbrace{adjP_0b_0}_{=\overline\lambda_0}=0
$$
(where $I$ is the identity matrix) to conclude that $\overline\lambda_0=0$. Since $rank(P_0)=n-1$ implies $rank(adj(P_0))=1$, the equation above also implies that not being on the \GB manifold is equivalent of
\begin{align}
\overline\lambda_0
\neq 0
\end{align}
We will use this identity later on.

The velocity $\dot q(t)$ is a continuous function during impact-free motion, hence $|\dot q|$ has a global maximum $\dot q_{max}$ over the closed interval $t\in[t_{init},0]$. 
In addition, $P$ is assumed to be Lipschitz with respect to $q$, which means that $det(P)$ is also Lipschitz continuous function of $q$. Let $\mathcal L$ denote a Lipschitz constant of $det(P)$. 
The two properties above can be combined with \eqref{eq:detP=0} to obtain
\begin{align}
|detP(q(t))|<\underbrace{det(P_0)}_0+\mathcal{L}|q(t)-q_0|\leq \mathcal{L}\dot q_{max}(-t)
\label{eq:detPbound}
\end{align}
which shows that the contact force \eqref{eq:lambda P} has a $1/x$ type singularity.

Next we examine how the tangential velocity of the contact points varies in response to the singular contact force. The tangential acceleration of the contact points can be expressed according to \eqref{eq:xdyn} and \eqref{eq:lambda P} as
\begin{align}
\ddot x=a+K\cdot det(P)^{-1}\overline\lambda
\end{align}
where the second term inherits the singularity at $t=0$. 
We can rescale this equation as
\begin{align}
det(P)\ddot x=a\cdot det(P)+K\overline\lambda
\end{align}
where all terms on the right-hand side of the equation are continuous functions of time, hence $det(P)\ddot x$ is also continuous. In the limit $t\rightarrow 0$, we have
\begin{align}
\lim_{t\rightarrow 0}det(P)\ddot x=K_0\overline\lambda_0
\end{align}
It is straightforward to prove that the vector $K_0\overline\lambda_0$ is not the zero vector, i.e. there is a scalar $1\leq j\leq n$ such that 
\begin{align}
\lim_{t\rightarrow 0} det(P)\ddot x_j\neq0
\end{align}
Due to the continuity of function $det(P)\ddot x_j$ 
we can choose a time $t_1$ sufficiently close to $0$ such that 
$det(P)\ddot x_j$ is bounded away from 0 with some bound $E$: 
\begin{align}
|det(P)\ddot x_j|>E>0
\end{align}
We rearrange this equation as
\begin{align}
|\ddot x_j|>E\cdot |det(P)^{-1}|
\end{align}
and by using \eqref{eq:detPbound}, we arrive to
\begin{align}
|\ddot x_j|>\frac{E}{\mathcal{L}\dot q_{max}}(-t)^{-1}
\end{align}

This bound involves a non-integrable singularity, which means that the variation of $\dot x$ in the time interval $[t_1,0]$ is unbounded. In other words, for any positive sliding velocity $\dot x_j(t_1)$, either $\dot x_j = 0$ is reached for some $t<t_0$, or $\dot x_j$ grows unbounded. The first scenario means that a slip-stick transition occurs, whereas the second is impossible due to the bounded mechanical energy of the system. We conclude that the system may not slip until $t=0$.

\bibliographystyle{plainnat}
\bibliography{painleve}

\end{document}